\documentclass[11pt,a4paper]{article}\usepackage[]{graphicx}\usepackage[]{color}
\makeatletter
\def\maxwidth{ %
  \ifdim\Gin@nat@width>\linewidth
    \linewidth
  \else
    \Gin@nat@width
  \fi
}
\makeatother

\definecolor{fgcolor}{rgb}{0.345, 0.345, 0.345}

\usepackage{framed}
\makeatletter
 {\par\unskip\endMakeFramed%
 \at@end@of@kframe}
\makeatother

\definecolor{shadecolor}{rgb}{.97, .97, .97}
\definecolor{messagecolor}{rgb}{0, 0, 0}
\definecolor{warningcolor}{rgb}{1, 0, 1}
\definecolor{errorcolor}{rgb}{1, 0, 0}
\newenvironment{knitrout}{}{} 

\usepackage{alltt}
\usepackage[a4paper, total={6.5in, 10in}]{geometry}
\usepackage{parskip}
\usepackage{amsmath}

 \usepackage[authoryear]{natbib}
\usepackage[utf8]{inputenc}

\usepackage{afterpage}

\pdfpageattr {/Group << /S /Transparency /I true /CS /DeviceRGB>>}

\usepackage{xargs}                      
\usepackage[prependcaption,textsize=small]{todonotes}

\newcommandx{\unsure}[2][1=]{\todo[linecolor=red,backgroundcolor=red!25,bordercolor=red,#1]{#2}}
\newcommandx{\change}[2][1=]{\todo[linecolor=blue,backgroundcolor=blue!25,bordercolor=blue,#1]{#2}}
\newcommandx{\info}[2][1=]{\todo[linecolor=OliveGreen,backgroundcolor=OliveGreen!25,bordercolor=OliveGreen,#1]{#2}}
\newcommandx{\improvement}[2][1=]{\todo[linecolor=Plum,backgroundcolor=Plum!25,bordercolor=Plum,#1]{#2}}
\newcommandx{\thiswillnotshow}[2][1=]{\todo[disable,#1]{#2}}

\makeatletter
\renewcommand*{\@textcolor}[3]{%
  \protect\leavevmode
  \begingroup
    \color#1{#2}#3%
  \endgroup
}
\makeatother

\newcommand{\changed}[1]{{\textcolor{black}{#1}}}

\newcommand{\comment}[1]{}

\usepackage{xifthen}            
\usepackage{array}
\usepackage{amssymb}

\newcommand{\blanco}[1]{  } 

\newcommand{\latin}[1]{\textit{#1}}

\newcommand{\abk}[1]{\mbox{#1}\xdot}
\DeclareRobustCommand\xdot{\futurelet\token\Xdot}
\def\Xdot{\ifx\token\bgroup.\else\ifx\token\egroup.\else
  \ifx\token\/.\else\ifx\token\ .\else\ifx\token!.\else
  \ifx\token,.\else\ifx\token:.\else\ifx\token;.\else
  \ifx\token?.\else\ifx\token/.\else\ifx\token'.\else
  \ifx\token).\else\ifx\token-.\else\ifx\token+.\else
  \ifx\token~.\else
  \ifx\token.\else.\ \fi\fi\fi\fi\fi\fi\fi\fi\fi\fi\fi\fi\fi\fi\fi\fi}

\newcommand{\eg}{\abk{\latin{e.\,g}}}   
\newcommand{\ie}{\abk{\latin{i.\,e}}}

\newcolumntype{L}[1]{>{$}p{#1}<{$}} 
\newcolumntype{M}{>{$}l<{$}}    
\newcolumntype{N}{>{$}c<{$}}    

{\section*{Acknowledgments}}%
{}

\DeclareMathOperator{\Bin}{Bin} 
\DeclareMathOperator{\Nor}{N} 
\DeclareMathOperator{\MVN}{MVN} 
\DeclareMathOperator{\HN}{HN} 
\DeclareMathOperator{\LogitN}{Logit N} 
  
\DeclareMathOperator{\Be}{Be} 
\DeclareMathOperator{\E}{\mathbb{E}} 
\renewcommand{\P}{\operatorname{\mathsf{Pr}}} 
\DeclareMathOperator{\p}{p}     
\DeclareMathOperator{\B}{B}     
\renewcommand{\L}{\operatorname{L}}     

\newcommand{\ml}[2][]{
  \ifthenelse{\isempty{#1}}%
  {\widehat{#2}_{\scriptscriptstyle{ML}}}%
  {\widehat{#2}^{#1}_{\scriptscriptstyle{ML}}}
}
\newcommand{\simiid}{\mathrel{\overset{iid}{\thicksim}}} 

\newcommand{\given}{\,\vert\,} 

\newcommand{\partialv}[3][]{%
  \ifthenelse{\isempty{#1}}
  {\frac{\partial\,#2}{\partial\,#3}}
  {\frac{\partial^{#1} #2}{\partial\,#3^{#1}}} 
} 

\newcommand{\partials}[3][]{%
  \ifthenelse{\isempty{#1}}
  {\frac{d\,#2}{d\,#3}}
  {\frac{d^{#1} #2}{d\,#3^{#1}}}
} 

\newcommand{\dseps}[2][]{%
  \ifthenelse{\isempty{#1}}
  {\frac{d}{d#2}}
  {\frac{d^{#1}}{d#2^{#1}}}
}

\newcommand{\dsepv}[2][]{%
  \ifthenelse{\isempty{#1}}
  {\frac{\partial\,}{\partial\,#2}}
  {\frac{\partial^{#1}}{\partial\,#2^{#1}}}
}

\renewcommand{\d}{\mathop{}\!\mathrm{d}}

\newcommand{\Ind}[1]{\mathcal{I}_{#1}}
\newcommand{\Infu}[1]{\mathcal{I}\left\{#1\right\}}

\newcommand{\mailto}[2][]{%
  \ifthenelse{\isempty{#1}}%
  {\href{mailto:#2}{\nolinkurl{#2}}}%
  {\href{mailto:#2}{\nolinkurl{#1}}}%
}

\DeclareMathOperator*{\argmax}{arg\,max}

\renewcommand{\vec}[1]{\boldsymbol{#1}}

\newcommand{\ntt}{\normalfont\ttfamily}

\newcommand{\pkg}[1]{{\protect\ntt#1}}

\IfFileExists{upquote.sty}{\usepackage{upquote}}{}
\begin{document}

\title{Power Priors Based on Multiple Historical Studies \changed{for Binary Outcomes}}
\author{Isaac Gravestock and Leonhard Held* \\
Epidemiology, Biostatistics and Prevention Institute\\
University of Zurich\\
\small{*leonhard.held@uzh.ch}
}
\date{\today}
\maketitle


\begin{abstract}
Incorporating historical information into the design and analysis of a
new clinical trial has been the subject of much recent discussion.
For example, in the context of clinical trials of antibiotics for drug
resistant infections, where patients with specific infections can be
difficult to recruit, there is often only limited and heterogeneous
information available from the historical trials. To make the
best use of the combined information at hand, we consider an approach
based on the multiple power prior which allows the prior weight of
each historical study to be chosen adaptively by empirical Bayes. This
choice of weight has advantages in that it varies commensurably with
differences in the historical and current data and can choose weights
near 1 if the data from the corresponding historical study are similar
enough to the data from the current study.  Fully Bayesian approaches
are also considered. The methods are applied to data from antibiotics
trials. An analysis of the operating characteristics in a {binomial 
setting} shows that the proposed empirical Bayes adaptive method works
well, compared to several alternative approaches, including the
meta-analytic prior.

\end{abstract}
{\bf Key Words}:  {Clinical Trials; Empirical Bayes; Historical Controls; Operating Characteristics; Power Prior} 

\section{Introduction}

A growing body of literature examines the use of historical data to augment newly collected data in clinical trials where patients are difficult to recruit \citep{spiegelhalter2004bayesian}.
\citet{viele2014use} review a variety of methods for incorporating historical data, focusing on the case of a single historical study. 
If more than one relevant source of historical information is available, then all of these should be incorporated into the calculations, however not all of the methods can easily be extended to handle many sources.
\changed{
\citet{doi:10.1080/19466315.2016.1174149} describe the importance of using the available data in clinical trials and how this can be done appropriately.
\citet{Rosmalen2017} do a broad study of methods for incorporating multiple sources of historical data into a clinical trial analysis, looking at a time-to-event endpoint.
\citet{doi:10.1002/pst.1722} describes the problem of prior-data conflict which can occur when including historical data and describe approaches to prevent them.}
We look more closely at two of the most discussed methods for incorporating historical information: power priors \citep{ibrahim2015power} and meta-analytic priors \citep{neuenschwander2010summarizing}. 

The meta-analytic predictive prior (MAP) handles many studies by design. It is based on well known meta-analysis models, and it has been used in a number of published studies \citep{Hueber2012, Baeten2013}. 
One challenge in its use is in the careful specification of the prior on the between-study heterogeneity parameter, as this determines the strength of the prior. It has also been extended and made robust \citep{schmidli2014robust}. The model is appealing for its simplicity and familiarity, but the exchangeability assumption is not necessarily suitable in every situation and other more flexible models may be better suited to incorporating the heterogeneity of the historical studies into a prior.

The prior model we consider in more detail is the power prior \citep{ibrahim2015power}, which is based on the idea of down-weighting the likelihood of historical data. This down-weighting is done with parameter, $\delta$, between 0 and 1 as a power on the likelihood of the historical data. 
We look at the power prior for single studies, with conditional weights and with random weights, and then the extension of each to multiple studies. 
While the idea is simple, the formulation is complicated by the need to correctly normalise the prior when using random weights and this makes using multiple historical studies a challenge.
While the extension to multiple studies is not new, there has been limited research on its application and the choice of weight parameters and priors in the multiple historical trial setting.

For the case of a single historical study, \citet{ibrahim2000power}
defined what is now considered the conditional power prior (CPP):
$\p(\theta_\star \given x, \delta) \propto \L(\theta_\star;x, n)^{\delta} \p(\theta_\star)$,
where $\delta \in (0,1)$ is the prior
weight, $x$ is the result from the historical study with sample size $n$ and $\theta_\star$
is the parameter of interest in the new study.  To handle unknown weights,
\citet{duan2006evaluating} extended the power prior formulation to
include the necessary normalising factor \citep{neuenschwander2009note} and a prior on the study
weight, $\p(\delta)$. The normalised power prior (NPP) is the joint distribution of $\theta_\star$ and $\delta$, defined as the product of the normalised conditional distribution of $\theta_\star\given x,\delta$ and the prior distribution of $\delta$,
\begin{align}
\p(\theta_\star, \delta \given x) &= \p(\theta_\star \given x, \delta)\p(\delta)  \nonumber \\
&= \frac{1}{C(\delta, x,n)} \L(\theta_\star; x, n)^{\delta} \p(\theta_\star) \p(\delta),  \label{eq:single_npp_post}
\end{align}
where 
\[
C(\delta, x, n) = \int \L(\theta_\star; x, n)^{\delta} \p(\theta_\star) \d\theta_\star.
\]
The NPP posterior of $\theta_\star$ and $\delta$ is
\begin{align*}
 \p(\theta_\star, \delta \given x, x_\star) \propto \L(\theta_\star; x_\star, n_\star) \p(\theta_\star, \delta \given x),
 \end{align*}
where $x_\star$ denotes the result of the new study with sample size $n_\star$. 
\citet{Gravestock2016} examined approaches to choosing power priors in
the single study setting and proposed using an empirical Bayes type approach
as an alternative to fully Bayesian methodology. They studied the operating
characteristics and found both approaches performed well. 
\citet{hees2017blinded} further examine how the empirical Bayes approach performs in a trial design involving blinded sample size recalculation.

\changed{
Since we wish to construct priors based on the data from several previously
conducted studies, we need to fix some further notation. Let there be $H$ historical studies
with indices $\mathcal{H}=\{1,\ldots,H\}$.} As before, the new study for which we
require the prior is denoted with $\star$. For each historical study
$i\in \mathcal{H}$ there is a result $x_i$ (a realisation of a random
variable $X_i$) based on a study specific parameter $\theta_i$, and a
sample size $n_i$.  For the
new study these quantities are, respectively, $x_\star, \theta_\star,
n_\star$. For vectors of values we use bold face, \eg
$\vec{x}=(x_1,...,x_H)$.

The CPP can easily be extended to handle multiple historical studies, 
\[ \p(\theta_\star \given \vec{x}, \vec{\delta}) \propto \left\{ \prod_{i \in \mathcal{H}} \L(\theta_\star; x_i, n_i)^{\delta_i} \right\} \p(\theta_\star).\]
For the NPP, the extension to multiple studies has been briefly discussed by
\citet[Sec 4.2]{Duan2005}, who describes three possible methods.  We
find that only the first of these gives reasonable definitions of the
prior and posterior. The second is equal to the first when properly 
normalised and the third is mathematically problematic. The first is defined as
the normalised version of the multiple CPP multiplied by the prior on $\vec\delta$,
\begin{align}
\p(\theta_\star, \vec{\delta} \given \vec{x}) &=  \p(\theta_\star \given\vec{x}, \vec\delta)\p(\vec\delta) \nonumber \\
&= \frac{1}{C(\vec{\delta}, \vec{x}, \vec{n})} \prod_{i\in\mathcal{H}} \left\{ \L(\theta_\star; x_i, n_i)^{\delta_i} \right\} \p(\theta_\star) \p(\vec\delta),\label{eq:npp_D_prior}
\end{align}
where
\begin{align*}
C(\vec{\delta}, \vec{x}, \vec{n}) &= \int \prod_{i\in\mathcal{H}} \left\{ \L(\theta_\star; x_i, n_i)^{\delta_i}  \right\}\p(\theta_\star) \d\theta_\star. 
\end{align*}
Since the likelihoods are multiplied together and then normalised,
there is a dependence between the studies' data and parameters and
\changed{their $\delta_i$ values.}

There has been some criticism of the NPP
\eqref{eq:npp_D_prior} due to its computational difficulty. This is
because of two different integration problems associated with the
formulation. The first is related to the number of model parameters
which need to be integrated out to find the normalising constant
$C(\vec{\delta}, \vec{x}, \vec{n})$.  In regression models where there is a
parameter for each covariate, this can easily become
unworkable. \citet{Dejardin2013} and \citet{Rosmalen2017} propose methods to 
simplify or approximate these calculations.
In the simple case presented in this paper, we only have a single model
parameter $\theta_\star$, so this integral is not problematic.  The
second potentially problematic integral comes in the calculation of
the marginal prior distribution of $\theta_\star$ from
\eqref{eq:npp_D_prior}, which requires integration with respect to
$\vec\delta$ in the fully Bayesian approach. Numerical approaches are 
required for this integral and this is discussed further
in Section \ref{sec:priorModels}.

\citet[Ch 5]{spiegelhalter2004bayesian} \changed{provide} a framework to classify the different models used to construct priors based on historical data.
Accordingly, the MAP prior is categorised as ``exchangeable" because the historical and current data are related through a common parameter and random effects distribution.
The power prior is based on an ``equal but discounted" relationship, \ie there is a parameter common to all studies, but the evidence from the historical studies about this parameter should be discounted.

In this paper, we extend the \citet{Gravestock2016} approach and consider
how empirical Bayes (EB) and full Bayes (FB) approaches can be applied in the
multiple historical study setting.  We examine the models of the power prior
and MAP methods (Section \ref{sec:priorModels}) and
apply those to derive priors based on recent clinical trials in the context 
of antibiotic development (Section \ref{sec:antibiotics}). 
We study their frequentist operating characteristics (Section 
\ref{sec:operatingCharacteristics}) and conduct a simulation study (Section
\ref{sec:simulations}). We close with some discussion in Section \ref{sec:discussion}.

\section{Prior Models for Multiple Historical Studies}\label{sec:priorModels}

\subsection{Naïve Approaches to Multiple Power Priors}
Instead of using the multiple power prior formulation, one could attempt to directly apply the empirical Bayes methods for single \changed{studies} of \citet{Gravestock2016} to the multiple studies. We look at two possible naïve approaches.

\paragraph{Pooled}
\changed{
First, we could consider all of the historical studies as a single, indivisible source of data and apply the single method once. Therefore, we treat the the product of likelihoods of the historical studies in (\ref{eq:npp_D_prior}) as a single likelihood and estimate one shared $\delta$ (or equivalently $\delta_1=\cdots=\delta_H$ in the product).
In the binomial setting this is equivalent to summing the results from all of the historical trials into a single
pooled result, $x=\sum_{i \in \mathcal{H}} x_i$ and $n = \sum_{i \in \mathcal{H}} n_i$.
This approach is the least flexible, and is similar to the MAP prior in
that a single parameter is used to adjust the variance of all studies.
It is a rather strong assumption that all the historical trials should be treated as
a single large trial and should be down-weighted by the same factor. Treating them as a
single trial does not allow the model to adapt to the heterogeneity that exists within the
historical data, in this way it is similar to a fixed-effects model.}

 
\paragraph{Separate}
 
Another possibility is to consider the historical studies
independently, and estimate each $\delta_i$ separately, applying the 
single EB methodology $H$ times. 
By this method the weights are determined based on the difference between 
each historical study and the new study. These weights do not necessarily
adapt the combination of the historical data to the new data. The basis for this
method is to maintain the interpretation of the weights as in the
single setting, where each weight is a measure of the similarity of
the historical study to the new study.

\subsection{Full Bayes Multiple Power Priors}
In the fully Bayesian approach using (\ref{eq:npp_D_prior}), we specify prior distributions for all parameters, \ie, the weights $\vec\delta$, and the study parameter $\theta_\star$. Since we wish to construct the prior solely based on the historical data, \changed{we use an uninformative initial prior on $\theta_\star$, which for location parameters such as the binomial probability, we use a uniform distribution. }

The choice of prior for the weights is more interesting.
Since the weights are restricted to $(0,1)$, it is common in the single setting to specify a beta distribution for the prior. The obvious extension to the multiple setting is to have $\delta_i \simiid \Be(\alpha,\beta)$. A default choice for the parameters is $\alpha=\beta=1$. 
It also has been suggested to use smaller parameters \citep{dejardin2014}, \eg $\alpha=\beta=0.5$, such that the prior is still symmetrical, but has increased variance. This also has the interpretation of favouring either strong or weak borrowing.

Another alternative is to have a prior which does not assume independence between the weights. 
One simple approach is to introduce a fixed positive correlation between the $\delta_i$. 
\changed{We can do this by using a Gaussian copula to join the uniform marginal priors on $\delta_i$ \citep{joe2014copula}.} This is done by transforming the weights by the inverse standard normal cumulative distribution function, \ie $\Phi^{-1}(\delta_i)$, and then setting the prior on $\Phi^{-1}(\vec\delta)$ to be a multivariate normal distribution with mean $\vec{0}$ and covariance matrix with 1 on the main diagonal and all other values identical to $\rho$, $0\leq\rho<1$. That is 
\[ \left[ \begin{array}{c}
\Phi^{-1}(\delta_1) \\
\Phi^{-1}(\delta_2) \\
\vdots \\
\Phi^{-1}(\delta_H) \end{array} \right] \sim \MVN\left(\vec\mu=
\left[ \begin{array}{c}
0 \\
0 \\
\vdots \\
0 \end{array} \right],
\vec\Sigma = \left[ \begin{array}{cccc}
1 & \rho & \cdots & \rho  \\
\rho & 1 & \cdots & \rho  \\
\vdots & & \ddots& \vdots\\
\rho & \cdots & \rho & 1 \end{array} \right]
\right).\]
\changed{The correlation specified with fixed $\rho$ approximately corresponds to the correlation of the $\vec\delta$, \eg with $\rho=0.5$ giving $cor(\vec\delta)\approx 0.49$, and the correlation equal to 0 or 1 respectively for $\rho=0$ or 1.
That means when $\rho=0$, this prior is equal to the independent uniform prior for each $\delta_i$.
And at the other limit, $\rho=1$, the formulation is equivalent to pooling the results of all the historical studies and having a uniform prior on the single $\delta$.}

For inference on the study parameter $\theta_\star$, we calculate the marginal power prior density $\p(\theta_\star \given \vec{x})$ from (\ref{eq:npp_D_prior}).
We compute this integral numerically using Monte Carlo and Rao-Blackwellisation \citep{Gelfand1990}. 
\changed{That is, we take $J$ samples of $\vec\delta$  from its prior and then calculate the conditional density, $\p(\theta_\star\given\vec\delta^{(j)},\vec{x})$, for each sample. We found $J\approx 1000$ samples to be accurate and fast to calculate. 
We then average over the conditional densities, so the marginal is approximated by $\p(\theta_\star \given \vec{x}) = 1/J \sum_{j=1}^J \p(\theta_\star \given\vec\delta^{(j)},\vec{x})$.
In the binomial setting, the conditional densities are beta densities, so the average is effectively a mixture model with $J$ components.
This approach to constructing the density is implemented in the \pkg{StudyPrior}\citep{StudyPrior} package for \pkg{R} (as well as the other priors discussed in this paper).
}

%
%

\subsection{Combined Empirical Bayes Power Prior} \label{sec:PPEB}
Instead of the naïve applications of the single study approach, we can construct an empirical Bayes approach which considers the combination of the historical studies.
To choose the optimal weights for the data, we set $\vec\delta$ to its maximal marginal likelihood estimate. We can derive the marginal likelihood analytically in the binomial setting. If we assume a flat initial prior on $\theta_\star$, we have
\begin{align*}
C(\vec\delta, \vec{x}, \vec{n}) &= \int_0^1 \prod_{i \in \mathcal{H}} \left\{ \binom{n_i}{x_i}^{\delta_i} {\theta_\star}^{\delta_i x_i} (1-\theta_\star)^{\delta_i(n_i-x_i)} \right\} \d\theta_\star \\
&= \prod_{i \in \mathcal{H}} \left\{ \binom{n_i}{x_i}^{\delta_i} \right\} \int_0^1  {\theta_\star}^{\sum_{i \in \mathcal{H}} \delta_i x_i} (1-\theta_\star)^{\sum_{i \in \mathcal{H}} \delta_i(n_i-x_i)} \d\theta_\star \\
&= \prod_{i \in \mathcal{H}} \left\{ \binom{n_i}{x_i}^{\delta_i} \right\} \B\left(1+\sum_{i \in \mathcal{H}} \delta_i x_i,1+\sum_{i \in \mathcal{H}} \delta_i(n_i-x_i)\right).
\end{align*}
The last equality follows from the definition of the beta function. 
Therefore,
\begin{align}\notag
\p(\theta_\star,\vec\delta \given \vec{x}) &= \frac{1}{C(\vec\delta, \vec{x}, \vec{n})} \prod_{i \in \mathcal{H}} \left\{ \binom{n_i}{x_i}^{\delta_i} {\theta_\star}^{\delta_i x_i} (1-\theta_\star)^{\delta_i(n_i-x_i)} \right\} \p(\vec\delta)\\
\notag
&=\frac{1}{C(\vec\delta, \vec{x}, \vec{n})} \prod_{i \in \mathcal{H}} \left\{ \binom{n_i}{x_i}^{\delta_i} \right\}   {\theta_\star}^{\sum_{i \in \mathcal{H}} \delta_i x_i} (1-\theta_\star)^{\sum_{i \in \mathcal{H}} \delta_i(n_i-x_i)}  \p(\vec\delta) \\
\notag
&= \B\left(1+\sum_{i \in \mathcal{H}} \delta_i x_i,1+\sum_{i \in \mathcal{H}} \delta_i(n_i-x_i)\right)^{-1}  {\theta_\star}^{\sum_{i \in \mathcal{H}} \delta_i x_i} (1-\theta_\star)^{\sum_{i \in \mathcal{H}} \delta_i(n_i-x_i)}  \p(\vec\delta)\\
\label{eq:npp_D_prior_binom}
&= \p(\theta_\star \given \vec\delta, \vec{x}) \p(\vec\delta),
%
%
\end{align}
where $\theta_\star\given \vec\delta,\vec{x} \sim \Be(1+\sum_{i \in \mathcal{H}} \delta_i x_i,1+\sum_{i \in \mathcal{H}} \delta_i(n_i-x_i) )$. 
\comment{Removed theta_star in  Beta distribution.}
For empirical Bayes we require the likelihood of $\vec\delta$, which is equal to the integral of the likelihood $\L(\theta_\star; x_\star, n_\star)$ times the power prior \eqref{eq:npp_D_prior_binom} when the $\p(\vec\delta)=1$. Therefore the marginal likelihood is
\begin{align}\notag
\p(\vec\delta\given \vec{x}, x_\star) &\propto \int_0^1 L(\theta_\star; x_\star, n_\star) \p(\theta_\star, \vec\delta \given \vec{x}) \d\theta_\star \\
\notag
&= \int_0^1 \binom{x_\star}{n_\star} {\theta_\star}^{x_\star}{(1-\theta_\star)}^{n_\star-x_\star} \Be(\theta_\star; 1+\sum_{i \in \mathcal{H}} \delta_i x_i,1+\sum_{i \in \mathcal{H}} \delta_i(n_i-x_i) ) \d\theta_\star \\
\label{eq:npp_D_ml}
&= \binom{n_\star}{x_\star} \frac{ \B(x_\star +1+\sum_{i \in \mathcal{H}} \delta_i x_i, n_\star - x_\star + 1+\sum_{i \in \mathcal{H}} \delta_i(n_i-x_i) )}{\B(1+\sum_{i \in \mathcal{H}} \delta_i x_i,1+\sum_{i \in \mathcal{H}} \delta_i(n_i-x_i) )},
\end{align}
where the integral result is by the combination of beta and binomial densities integrating to give a beta-binomial density.

\changed{The empirical Bayes estimate is then $\vec{\hat\delta}_{\text{EB}}= \argmax_{\vec\delta} \p(\vec\delta \given \vec{x}, x_\star)$, with components denoted $\hat\delta_i$.}
As \eqref{eq:npp_D_ml} is a beta-binomial density, it is known that there is no analytical form for the ML estimates, so we use a numerical optimisation method to find the maximising values. Additionally we have the constraint of $\vec\delta \in [0,1]^H$.
\citet{BertoliBarsotti1994} sets out conditions for the existence of the MLE for a beta-binomial model, which are not satisfied with only a single $x_\star$ and $n_\star$. This is easily seen with the mean/size parameterisation, where $\mu=\alpha/(\alpha+\beta)$ and $M=\alpha+\beta$. Consider Figure \ref{fig:betabinomialML}, which shows the likelihood surface for $n_\star=100$, $x_\star=65$. The likelihood increases as $M\to\infty$. 
However, with the additional constraint on $\vec\delta$, the parameter space of $\{\mu,M\}$ is restricted, so a MLE can be found.
For data $\vec{x}/\vec{n}=40/90,50/80,60/90$, the region is shown in red in Figure \ref{fig:betabinomialML}. Finding the maximum likelihood estimate  of $\vec\delta$ (the red dot in Figure \ref{fig:betabinomialML}) is easy because the likelihood surface is smooth and the estimate always lies on the boundary of this region.

\begin{knitrout}
\definecolor{shadecolor}{rgb}{0.969, 0.969, 0.969}\color{fgcolor}\begin{figure}

{\centering \includegraphics[width=\maxwidth]{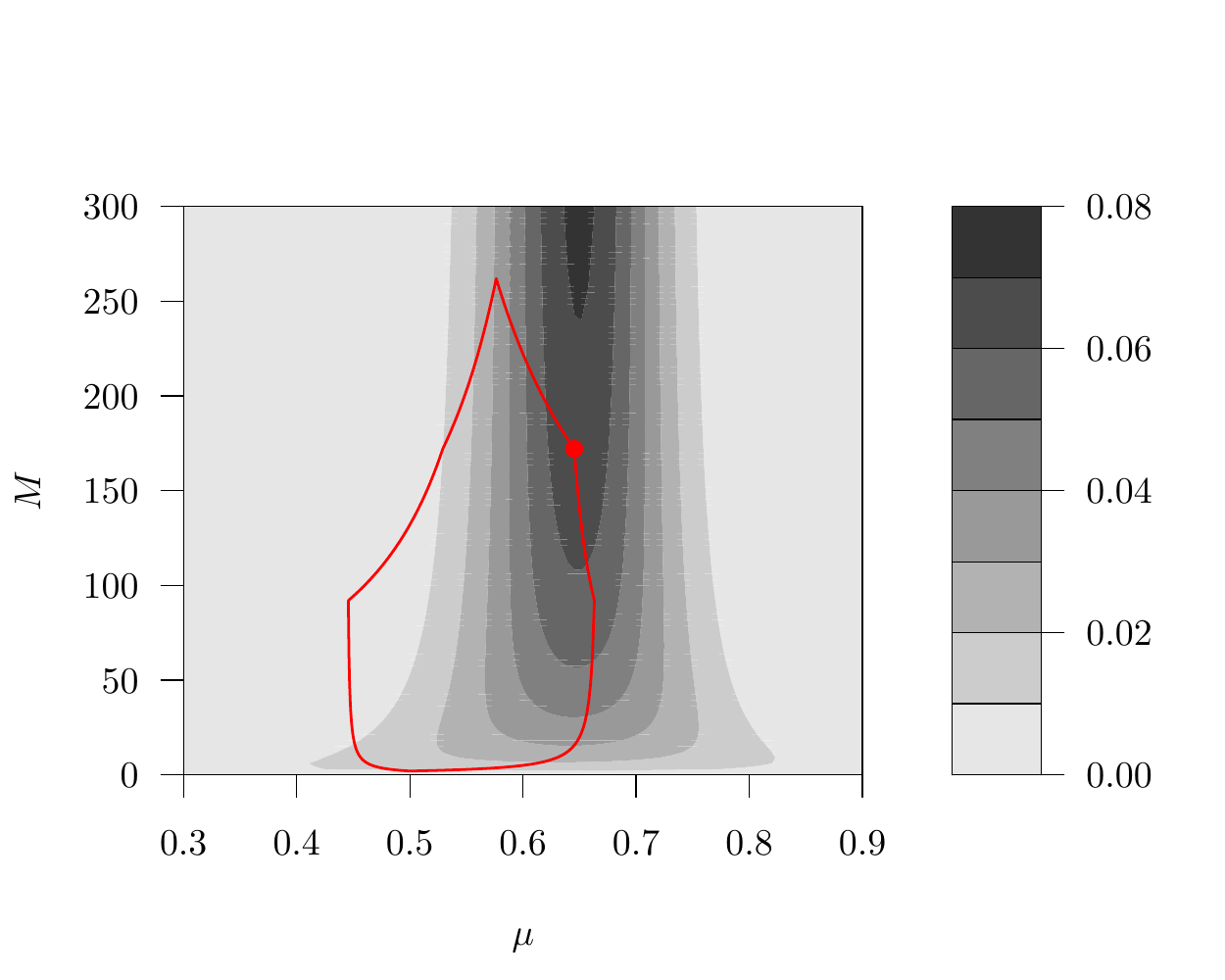} 

}

\caption[Likelihood surface of beta-binomial for $x=65$, $n=100$ and varying $\mu$ and $M$]{Likelihood surface of beta-binomial for $x=65$, $n=100$ and varying $\mu$ and $M$. The red line indicates the region which fulfills the additional constraint on $\vec\delta$, the red dot is the corresponding MLE}\label{fig:betabinomialML}
\end{figure}

\end{knitrout}

\subsection{Meta-Analytic Predictive Prior}
For comparison, we also consider the meta-analytic predictive prior (MAP)\citep{neuenschwander2010summarizing}, which is based on a random effects model for the historical and current study parameters. We assume that the random effect distribution is normal, so we use a logit transformation to get the probability parameters for the binomial distributions. The underlying parameters are denoted $\mu$ and $\tau$. Therefore the model is
\begin{align*}
\theta_1,...,\theta_H, \theta_\star \simiid& \LogitN(\mu, \tau^2)\\
 X_1 \given \theta_1 \sim& \Bin(\theta_1; n_1)\\
 \vdots&  \\
 X_H \given \theta_H \sim& \Bin(\theta_H; n_H)\\
 X_\star \given \theta_\star \sim& \Bin(\theta_\star; n_\star).
\end{align*}
To complete the model specification we need to choose fixed values or hyper-priors for $\mu$ and $\tau$. 
Since we want the historical studies to define the prior, \changed{we use a weakly informative prior for the location parameter, $\mu \sim \Nor(0,\pi^2/3)$, approximating a uniform prior on the probability scale.}


For a fully Bayesian MAP, we also specify a hyper-prior for $\tau$. It is recommended to use a prior that puts most of the mass on small values but still allows for the possibility of large variation. Suggested distributions for the prior are half-normal or $t$ distributions \citep[Ch 8]{spiegelhalter2004bayesian}, with parameters chosen to give a small probability of large between-study heterogeneity.
Following \citet{neuenschwander2010summarizing}, we use a half-normal prior, \ie $\tau \sim \HN(1)$. 
\comment{reference choice of prior}

The prior for the new study parameter is the marginal predictive distribution of the current study parameter based on the historical studies, $\theta_\star \given x_1, ...,x_H$.
This formulation assumes exchangeability between all of the studies. Due to this, the two stage calculation is equivalent to conducting a single meta-analysis using all of the studies' results \citep{schmidli2014robust}. 

\paragraph{Robustified Extension}
\citet{schmidli2014robust} suggest robustifying the MAP prior by approximating the prior density with a mixture of densities, and then adding an additional vague component, which gives the prior some mass across the whole parameter space of $\theta$. In this setting we use beta densities as they are conjugate with the binomial likelihood. We use numerical optimisation methods to select the optimal mixture weights and parameters to approximate the MAP. Then the vague component, $\Be(1,1)$, is added so that it has a 10\% weight in the final mixture. As this procedure only applies to the predictive distribution of the new study parameter, the current and historical studies are not exchangeable in the robust model.

%
%
\changed{
\subsection{Comparison Using Normal Likelihoods}\label{sec:normal}
By examining the priors discussed in this paper using normal models we can get an better understanding of the similarities and differences between the priors and their assumptions.
As noted in Section 1, there are a variety of possible models that could be used to construct a prior based on historical data, as laid out by \citet{spiegelhalter2004bayesian}, each with different assumptions about the differences in the historical and current data. \citet{chen2006relationship} describe how under certain conditions these different models can be equivalent.
}

\paragraph{Power Prior}
\changed{ The power prior is an equal but discounted model, which
  assumes the historical studies $\vec{X} = X_1, \ldots, X_H$ share the same
  underlying parameter $\theta_\star$, but some studies give better
  evidence for its estimation than others, which is quantified with
  the study specific power parameters. Generally we assume that the
  new study $X_\star$ provides the best evidence for the parameter of
  interest and the historical studies are discounted to provide only a
  fraction of evidence.  When using a normal likelihood, the power
  parameters $\delta_1,\ldots,\delta_H$ in the normalised power prior
  are divisors of the variances $\sigma^2_1,\ldots,\sigma^2_H$ of the
  historical studies. They can be be fixed or random parameters, here
  we assume they are fixed. See Appendix A for derivation of
  distribution of the normal power prior for random $\vec\delta$. We
  also assume a flat prior on $\theta_\star$ throughout this
  section. We can describe the normal power prior model with the
  following distributions: }

\changed{
\begin{align*}
 X_1\given \theta_\star \sim& \Nor(\theta_\star; \sigma^2_1/\delta_1)\\
 \vdots&  \\
 X_H\given \theta_\star  \sim& \Nor(\theta_\star; \sigma^2_H/\delta_H)\\
 X_\star\given \theta_\star   \sim& \Nor(\theta_\star; \sigma^2_\star)
\end{align*}
This implies that the power prior for $\theta_\star$ based on the historical data is
\[  \theta_\star \given \vec{X} = \vec{x} \sim \Nor\left(\frac{\sum_i x_i \delta_i/\sigma_i^2}{\sum \delta_i/\sigma_i^2}, \frac{1}{\sum \delta_i/\sigma_i^2} \right).  \]
Since $\delta_i \in (0,1]$, the modified variance term is larger than or equal to the observed variance of the study, \ie $\sigma_i^2/\delta_i \geq \sigma_i^2$.
It would of course be possible to express this inflated variance additively, \ie $\sigma_i^2/\delta_i = \sigma_i^2 + \tau_i^2$, where $\tau_i^2 = \sigma^2_i (1/\delta_i-1) \geq 0$. Rewriting the model using the this additive variance inflation gives
\begin{align*}
 X_1  \sim& \Nor(\theta_\star; \sigma^2_1 + \tau_1^2)\\
 \vdots&  \\
 X_H \sim& \Nor(\theta_\star; \sigma^2_H + \tau_H^2)\\
 X_\star  \sim& \Nor(\theta_\star; \sigma^2_\star),
\end{align*}
}

\changed{
This is then equivalent to a potential bias model \citep{pocock1976combination}, \citep[Section 5.4]{spiegelhalter2004bayesian}, where each historical study has a bias $b_i \sim \Nor(0, \tau_i^2)$ and the study-specific parameters are $\theta_i = \theta_\star+b_i$. Therefore the studies $1,..,H$ are not exchangeable with the new study $\star$.
The full potential bias model is
\begin{align*}
b_1 \sim& \Nor(0, \tau_1^2)\\
\vdots&  \\
b_H \sim& \Nor(0, \tau_H^2)\\
X_1\given b_1, \theta_\star  \sim& \Nor(\theta_\star + b_1; \sigma^2_1)\\
\vdots&  \\
X_H\given b_H, \theta_\star \sim& \Nor(\theta_\star+ b_2; \sigma^2_H)\\
X_\star \given \theta_\star \sim& \Nor(\theta_\star; \sigma^2_\star),
\end{align*}
}
\changed{
and, correspondingly, the power prior distribution can be written as 
\[  \theta_\star \given \vec{X} = \vec{x} \sim \Nor\left(\frac{\sum_i x_i/(\sigma_i^2 +\tau_i^2)}{\sum 1/(\sigma_i^2 +\tau_i^2)}, \frac{1}{\sum 1/(\sigma_i^2 +\tau_i^2)} \right).  \]
}
\changed{
\paragraph{MAP}
The MAP prior is built upon the exchangeability assumption, where each study has its own parameter of interest $\theta_i$ which are identically distributed around some underlying mean parameter, $\mu$,
\[\theta_1,...,\theta_H, \theta_\star \simiid \Nor(\mu, \tau^2).\]
The study data are then distributed around each of these as follows,
\begin{align*}
 X_1 \given \theta_1  \sim& \Nor(\theta_1; \sigma^2_1)\\
 \vdots&  \\
 X_H \given \theta_H  \sim& \Nor(\theta_H; \sigma^2_H)\\
 X_\star \given \theta_\star  \sim& \Nor(\theta_\star; \sigma^2_\star).
\end{align*}
Assuming a flat prior on $\mu$, the MAP prior for the parameter of the new study is then
\[  \theta_\star \given \vec{X} \sim \Nor\left(\frac{\sum_i x_i/(\sigma_i^2 +\tau^2)}{\sum 1/(\sigma_i^2 +\tau^2)}, \frac{1}{\sum 1/(\sigma_i^2 +\tau^2)} + \tau^2 \right).  \]
}
\changed{
\paragraph{Comparison}
The obvious difference between the two models is the additional variance term in MAP due to the difference in underlying parameter ($\mu$ versus $\theta_\star$). 
We can unify the models by rewriting the potential bias model to be centred around $\mu$, like the MAP. 
Therefore $\theta_i = \mu + b_i$ and now also $\theta_\star = \mu + b_\star$. It follows then that we also have a $\tau_\star^2$. 
Then the MAP has $\tau_1^2 = \cdots = \tau_H^2 = \tau_\star^2$, while the potential bias model has unconstrained $\tau_i^2$ but $\tau_\star^2 = 0$ which effectively makes $\mu = \theta_\star$.
With this unified notation, the difference between the models becomes clear because the 
exchangeability assumption of studies in the MAP model is not present in the power prior model. 
In principle the MAP model could
allow for subgroups of studies, \eg high quality and low quality, with different heterogeneity variances, say, $\tau_A^2$ and $\tau_B^2$, but then we must also carefully consider what heterogeneity exists between $\theta_\star$ and $\mu$.
}


\section{Clinical Example}\label{sec:antibiotics}

\begin{figure}[htbp]
 \includegraphics[width=\textwidth]{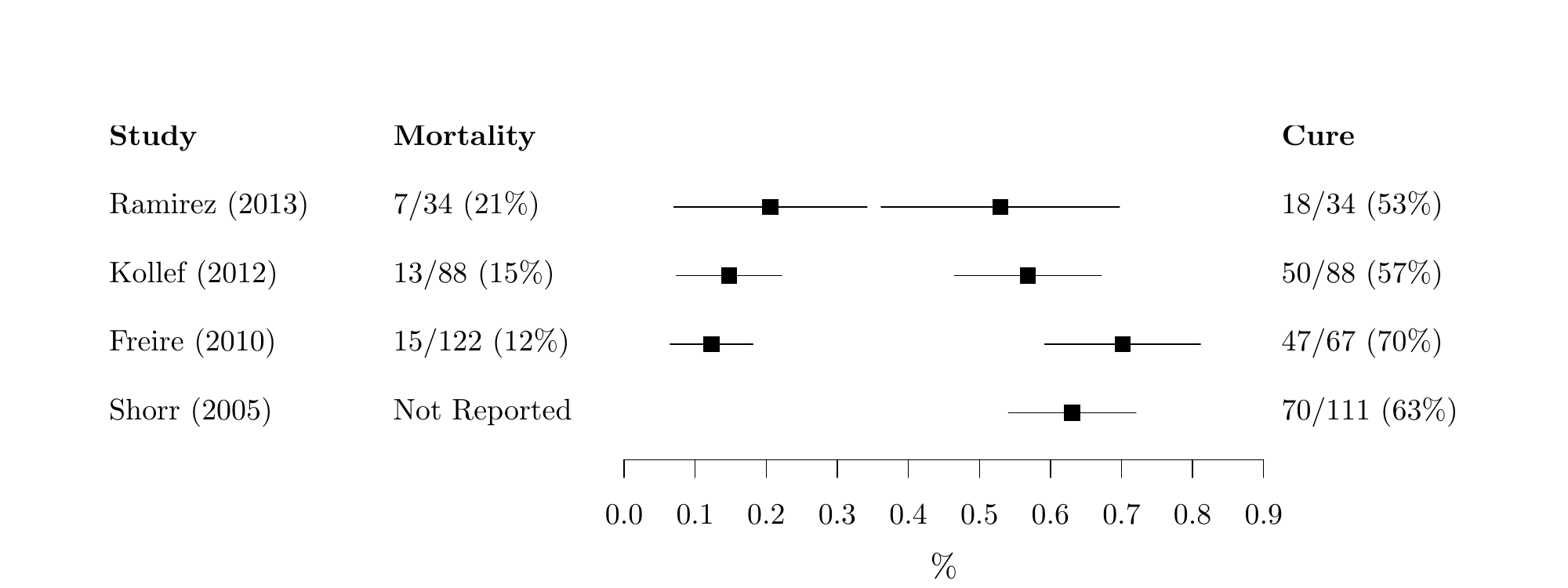}
\caption{\changed{Historical studies which could be used to construct a prior. The all cause mortality is shown on the left and clinical cure on the right of the forest plot.}}
\label{fig:Forestplot}
\end{figure}

To demonstrate how these methods could be applied, we consider constructing a prior for the cure rate in a clinical trial for treatment of ventilator-assisted pneumonia. These studies were found in recent meta-analysis of antibiotic therapies \citep{Cochrane2016} for ventilated pneumonia patients. Each of the chosen studies had a similar control arm treatment of combined imipenem/cilastatin. Figure \ref{fig:Forestplot} shows the results of three historical trials for two outcomes.

\changed{
Based on a fourth study, Ramirez (2013)\citep{ramirez2013randomized}, we demonstrate how these priors can be used.
The Ramirez trial planned to enroll 70 patients in the imipenem/cilastatin control arm, but was terminated due to difficulty in enrolling patients, having recruited only 34 patients in the control arm.
An informative prior might have helped reduce the number of patients required to get a suitably precise estimate of the cure rate of the treatment, and thus have allowed the trial to continue and be able to compare different treatments.
Alternatively, an analysis with the limited collected data combined with a prior based on historical data might allow some inference to be made.
}

\begin{table}[ht]
\centering
\begin{tabular}{lrrr|rrr}
  \hline
   & \multicolumn{3}{c}{{Cure}} & \multicolumn{3}{c}{{Mortality}}\\
 \cline{2-4}\cline{5-7}
 {Study} & {Sep} & {Comb} & {Pool} & {Sep} & {Comb} & {Pool}\\
 \hline
Kollef (2012) & 1.00 & 1.00 & 0.25 & 0.22 & 0.00 & 0.21 \\ 
  Freire (2010) & 0.12 & 0.00 & 0.25 & 1.00 & 1.00 & 0.21 \\ 
  Shorr (2005) & 0.51 & 0.00 & 0.25 &  &  &  \\ 
   \hline
\end{tabular}
\caption{Empirical Bayes $\vec{\hat{\delta}}$ values based on the limited data from the Ramirez trial for mortality (7/34) and clinical cure (18/34).} 
\label{tab:estimates}
\end{table}

Table \ref{tab:estimates} shows the estimates of $\vec\delta$ for the Ramirez trial using the methods in Section \ref{sec:PPEB} based on the small number of patients. Note the considerable differences in estimates between the different methods. The difference between Combined and Separate for the mortality outcome is especially large. This is due to the naivety of the Separate approach which chooses each $\delta_i$ independently, and, \changed{in this case, gives larger weights to all studies as each of them is similar to the Ramirez result.}
Conversely, the Combined approach only gives weight to the trial with the nearest result as including the others would further bias the posterior distribution away from the current data, and increase prior-data conflict.

\begin{figure}[htbp]
\begin{minipage}{.5\textwidth} 
 \includegraphics[width=\textwidth]{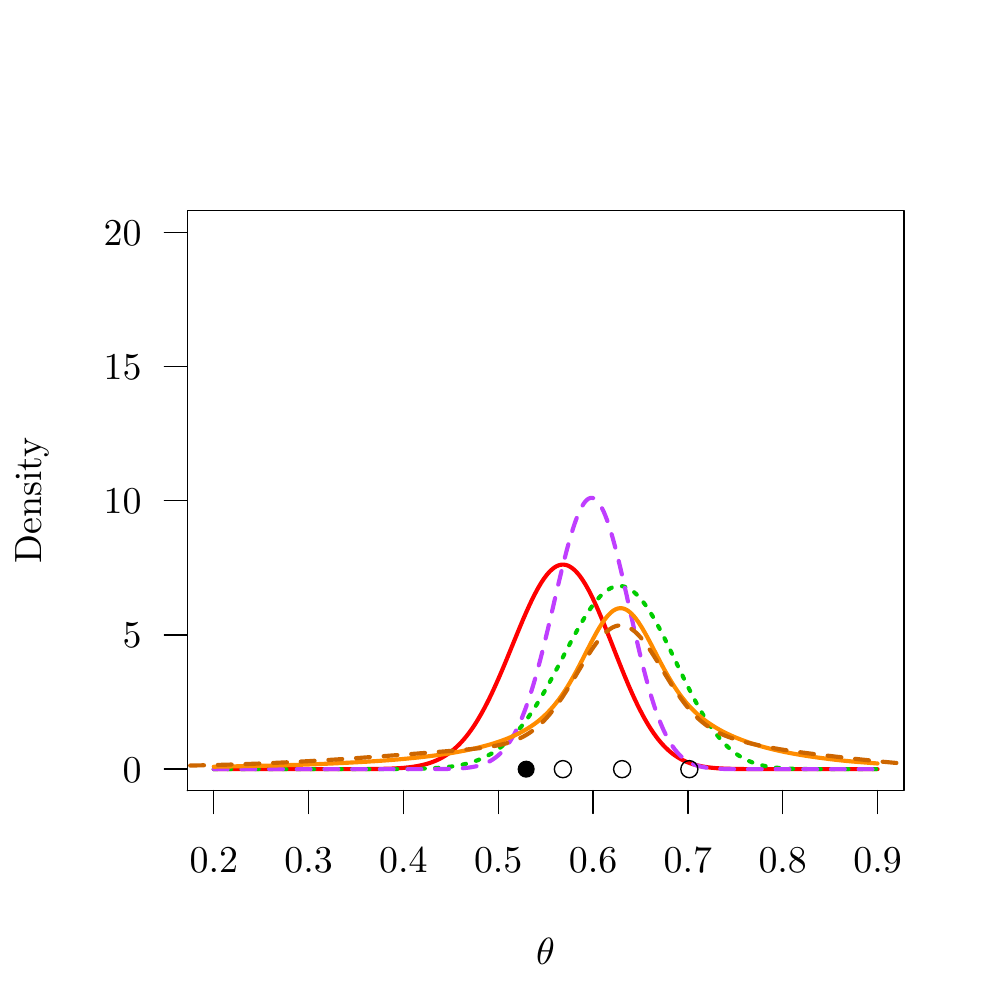}
 \centering (a) Clinical cure
 \end{minipage}
\hfill
\begin{minipage}{.5\textwidth}
 \includegraphics[width=\textwidth]{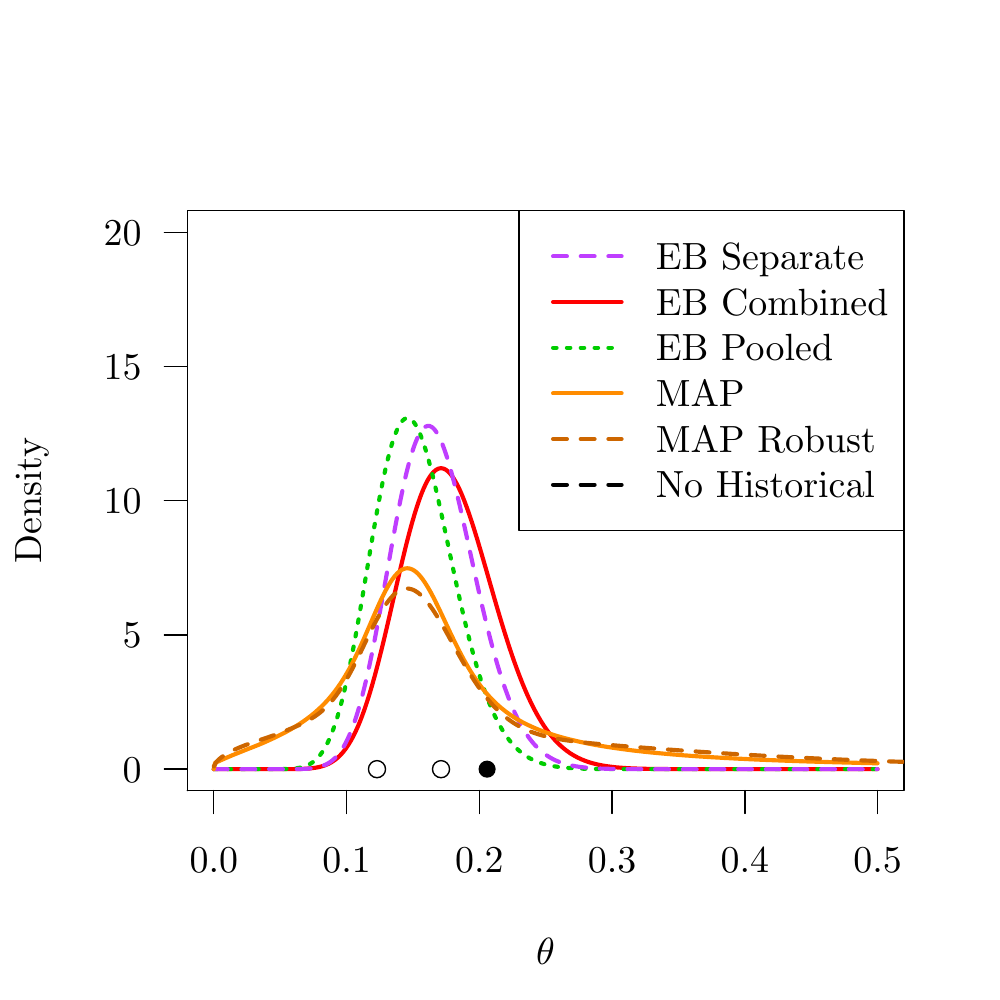}
 \centering (b) All cause mortality
\end{minipage}

\begin{minipage}{.5\textwidth} 
 \includegraphics[width=\textwidth]{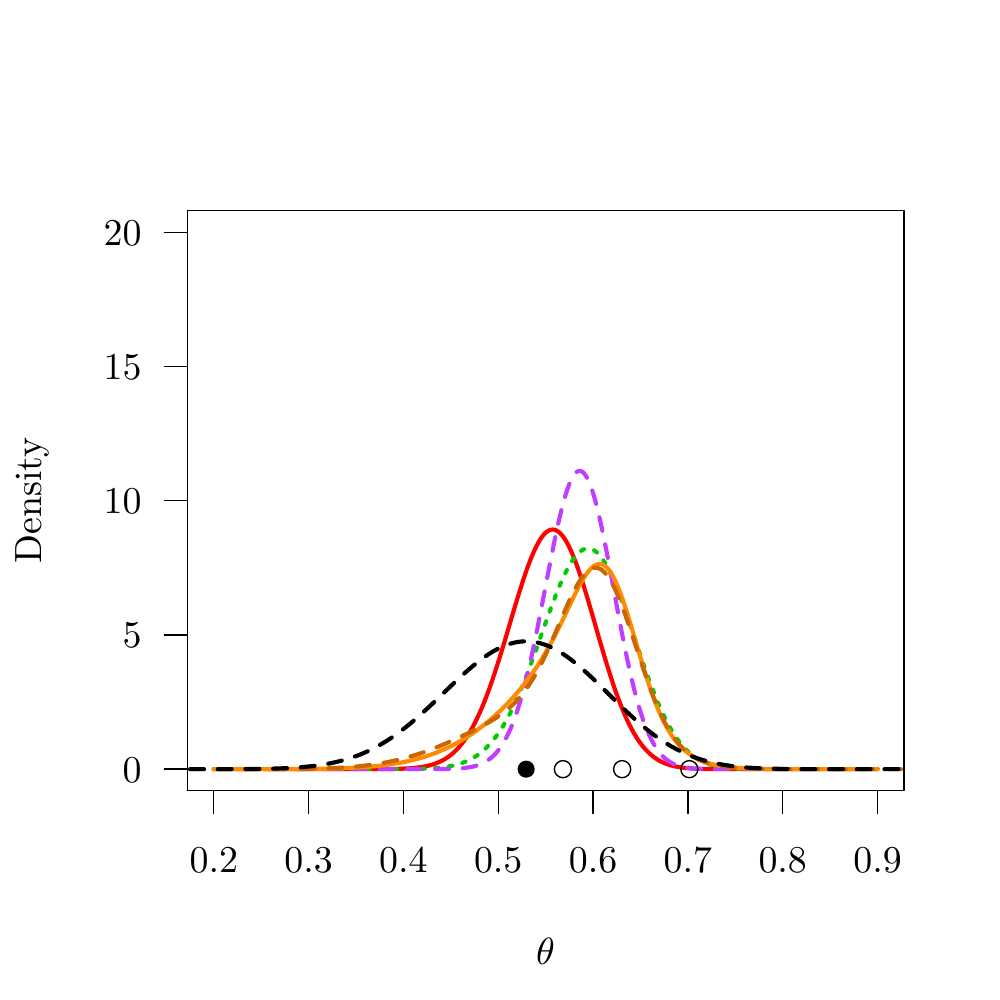}
 \centering (a) Clinical cure
 \end{minipage}
\hfill
\begin{minipage}{.5\textwidth}
 \includegraphics[width=\textwidth]{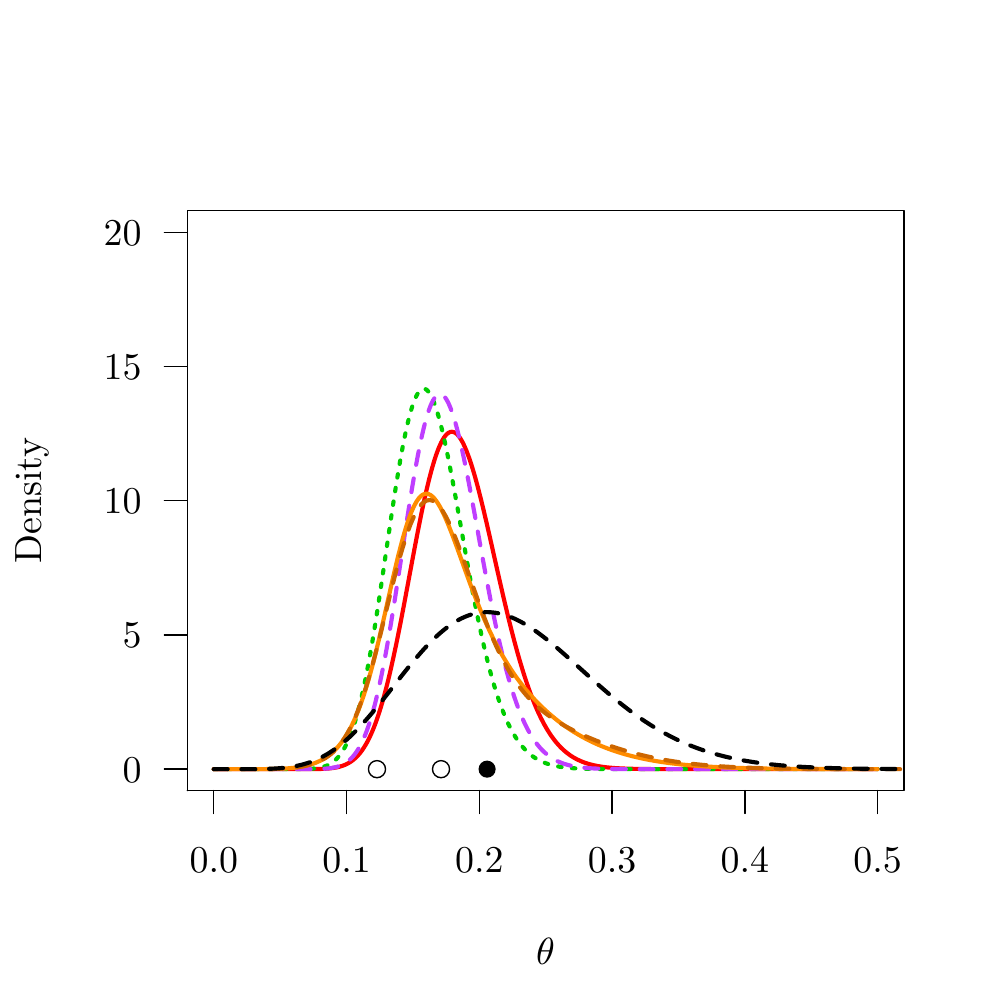}
 \centering (b) All cause mortality
\end{minipage}
\caption{Priors (above) and posteriors (below) of the EB Separate, Combined, and Pooled approaches, as well as MAP and MAP Robust compared with using no historical data. Open circles represent the historical data and the closed circles are the current data.}
\label{fig:Posteriors}
\end{figure}

\changed{Figure \ref{fig:Posteriors} shows the prior and posterior densities} resulting from these estimates as well as the MAP prior. 
As expected from the estimates in Table \ref{tab:estimates}, the Separate posterior is the narrowest for both outcomes because it incorporates the most historical data.
For the clinical cure outcome, the Pooled posterior incorporates very little historical data, so the posterior is most similar to that with no prior data. The Combined posterior is less extreme than either of the power prior approaches. 
The two MAP posteriors are quite similar. They have a mode centered over the historical data, but are skewed to have a fat tail covering the current data. The robust version is slightly more skewed towards the current data than the non-robust version, demonstrating the desired adaptiveness of the method.

\begin{figure}[htbp]
\begin{minipage}{0.32\textwidth} 
 \includegraphics[width=\textwidth]{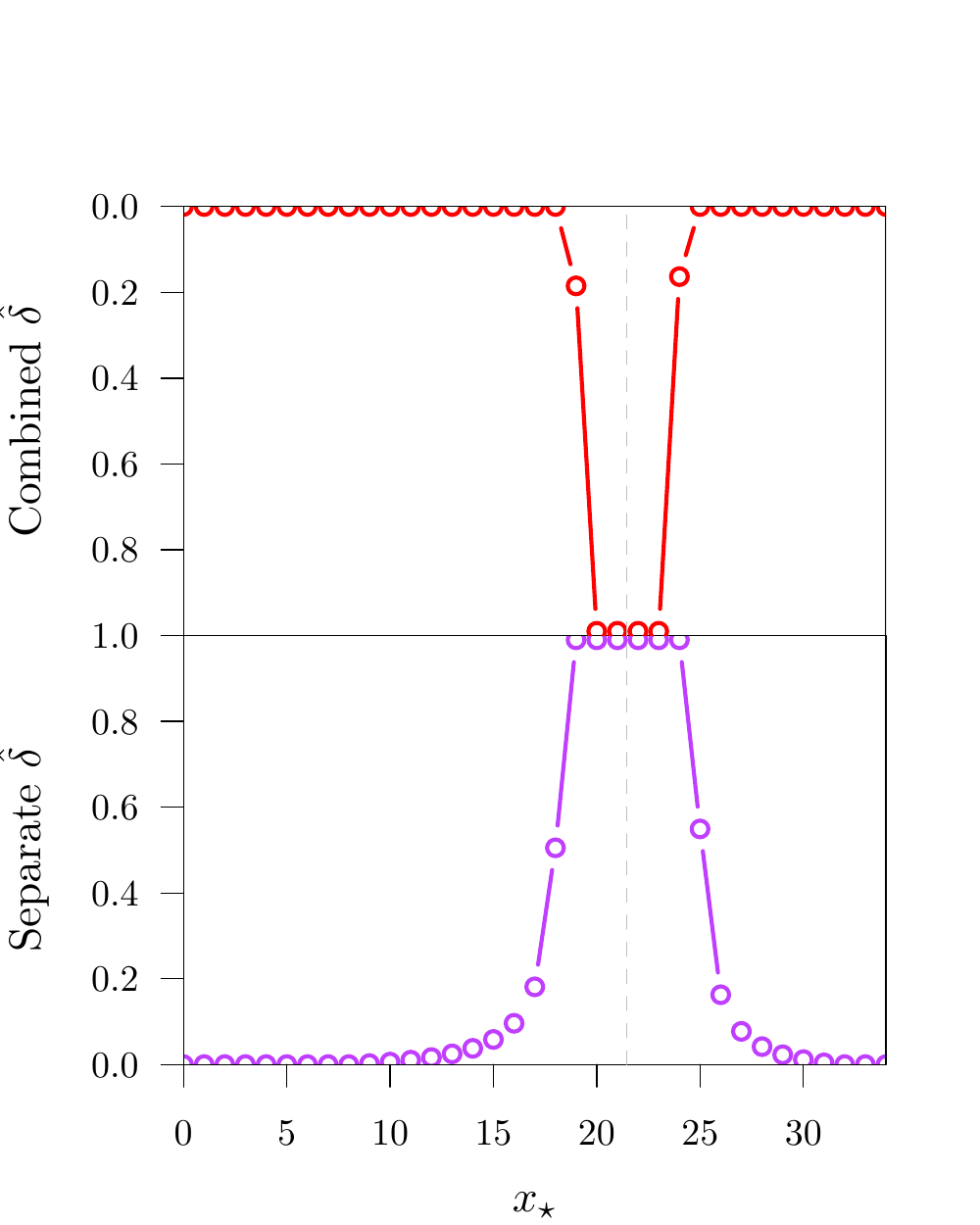}
 \centering (a) $\hat\delta$ for Shorr
 \end{minipage}
\hfill
\begin{minipage}{0.32\textwidth}
 \includegraphics[width=\textwidth]{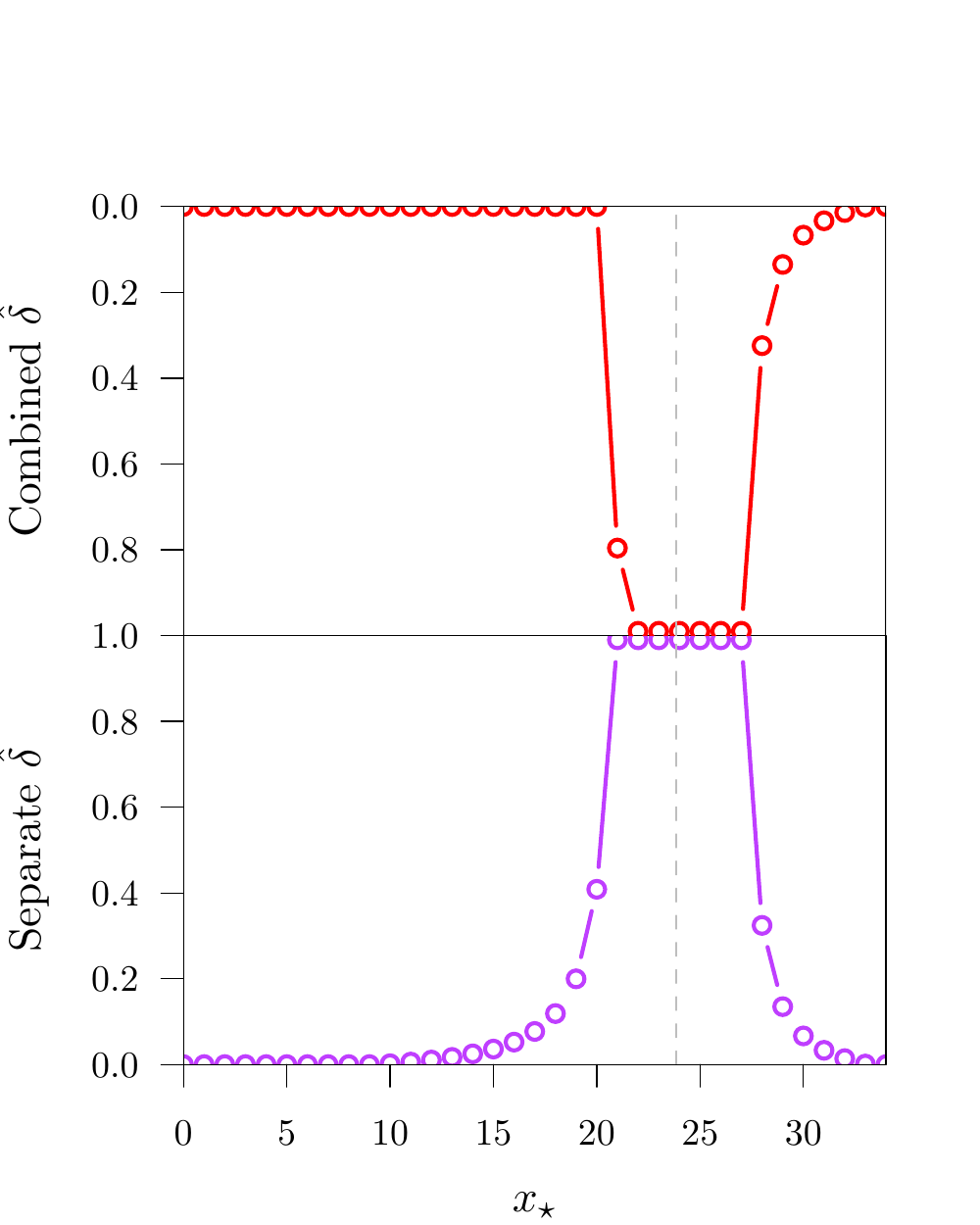}
 \centering (b) $\hat\delta$ for Freire
\end{minipage}
\hfill
\begin{minipage}{0.32\textwidth} 
 \includegraphics[width=\textwidth]{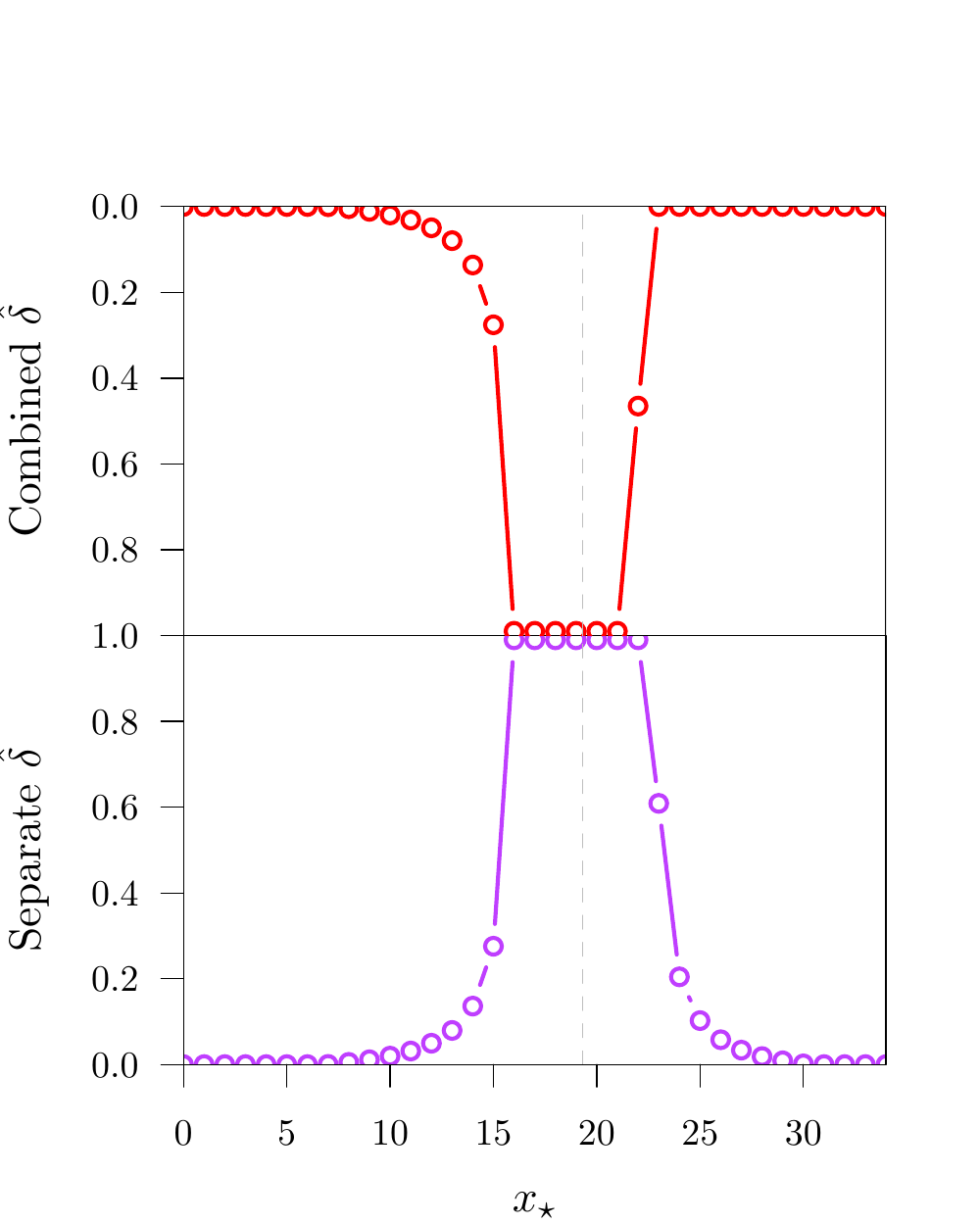}
 \centering (c) $\hat\delta$ for Kollef
\end{minipage}
\caption{Empirical Bayes estimates of $\delta_i$ for $x_\star$ for the clinical cure outcomes of the Shorr, Freire and Kollef studies based on a current study size $n_\star=34$. The dashed lines show where $x_\star/n_\star= x_i/n_i$ of the historical study.}
\label{fig:EBdelta2}
\end{figure}

To see the differences in $\vec\delta$ between methods, Figure \ref{fig:EBdelta2} compares the $\delta_i$ estimates of the Combined and Separate approaches for each historical study for the mortality outcome. Here we see that the Combined method borrows over a narrower range than the Separate. This is not generally true, however. If the results from the historical studies were more different, but were above and below the current study's result, then the range of borrowing for each trial would increase for Combined but not for Separate.
\changed{The estimation methods tend to choose either large or small values for $\delta_i$, with values intermediate only occurring in a relatively a narrow range of $x_\star$ values. This is the result of the method maximizing the marginal likelihood of $\vec\delta$ and only including studies which make the most likely prior for the new data. As a consequence of this somewhat binary behaviour, the EB power prior might then be compared to the `test-then-pool' method described by \citet{viele2014use}, which tests similarity between the current and historical study to determine if the data should be pooled. The advantage of the EB method is that it requires no specification of a cut-off level for the test and can flexibly choose different combinations of studies to construct a suitable prior. We are not aware of an extension of the `test-then-pool' methodology to multiple studies, and but the procedure could applied to each historical study individually in the same fashion as EB Separate.}

%

\section{Operating Characteristics}\label{sec:operatingCharacteristics}



%



For new statistical methods to be accepted for use in regulated environments they need to be thoroughly investigated.
Unlike frequentist methods, whose operating characteristics are often well defined by construction, Bayesian methods are more flexible and so simulation studies must be used to determine how they perform. Although Bayesian methods do not rely on the frequentist paradigm of repeated testing, it is still useful to test the frequentist operating characteristics of Bayesian methods \citep{Rubin1984, Grieve2016}. It is important to know how the inclusion of a certain prior may influence the outcome of an analysis and therefore how the study and testing procedure should be designed.
Similar to \citet{Gravestock2016}, we examine the prior sample size of the prior, the mean squared error of the posterior mean, the power, the type I error and the pre-experimental rejection ratio. 

\paragraph{Prior sample size}
To examine the influence of the prior, we consider its contribution to the posterior in terms of sample size and for the calculation of sample sizes it is useful to know how many patients worth of data the prior contributes.
To get the prior sample size, we calculate the sample size of the posterior and subtract $n_\star$.
We base these calculations on a beta prior, which due to conjugacy, means we can interpret the sum of the prior parameters as a sample size.

\changed{
The EB power prior methods in the binomial setting give a prior which is a beta distribution, so the posterior sample size is simply the prior sample size plus $n_\star$, so we can calculate the prior sample size by  $\sum_i \delta_i n_i$.
The FB power priors are calculated as the a of average $J$ beta densities, so we can approximate the prior sample size by $\sum_j^J \vec\delta^{(j)}\cdot \vec{n} - n_\star$. 
For priors which are not beta distributions, we approximate the densities with a mixture of conjugate distributions, in the binomial setting beta distributions, as in \citet{schmidli2014robust}. See Appendix B for more details.
We can approximate the average sample size of the posterior using the weighted sum of the beta parameters, similar to the FB power priors. 
The prior sample size is then
$\sum_j^J w_j' (\alpha_j' +\beta_j' )  - n_\star$.
This calculation of sample size is not exact, as there is an small inconsistency between the estimated sample sizes for mixtures with uniform components \ie $\Be(1,1)$ or components summing to uniform $0.5\Be(1,2) + 0.5\Be(2,1)$ giving sample size 2 or 3, respectively.
The impact of this inconsistency on our results is small, as the largest possible difference in sample size between combinations is only 1 patient.
}

\paragraph{Mean Squared Error}
\changed{
To understand how the prior affects the estimate of $\theta_\star$, we look at the mean squared error (MSE) of the posterior mean, $\hat\theta_\star= \int \theta_\star \p(\theta_\star \given \vec{x}, x_\star) \d \theta_\star $.
The MSE is the squared difference between the true parameter value, which we denote $\theta$ and the  expected value of estimate over all possible outcomes of the new study. For a given true parameter, it is defined as
\begin{align*}
\text{MSE}(\theta) = (\E\hat\theta_\star-\theta)^2
=  \sum_{x_\star=0}^{n_\star} \p(x_\star \given \theta) (\hat\theta_\star - \theta)^2.
\end{align*}
The MSE measures the quality of the posterior distribution as a predictive distribution of the true mean. Large values indicate either a strong bias or large variance and thus poor or uncertain estimation of the true mean.
}

\paragraph{Power}
The power of a study can be increased with the use of additional information. However, \citet{Cuffe2011} explains that including historical data may also lead to a decrease in power where the prior conflicts with the data. It is important that the prior does not negatively affect the feasibility of the study.
Therefore care must be taken that the power is not reduced dramatically by a conflicting prior.
\changed{To measure power, we calculate the probability of detecting a difference in parameter between the control arm and a treatment arm. Therefore we introduce the treatment arm with data $x_T$ from a binomial distribution with parameter $\theta_T$ and size $n_T = n_\star$.}
We use a Bayesian test where a difference is declared when the posterior probability of the treatment arm parameter being larger than that of the control arm is 0.975, \ie
\begin{align}
\Ind{x_\star,x_T} &= \Infu{\P(\theta_\star < \theta_T \given x_\star, x_T) >0.975} \nonumber \\
&= \Infu{\int_0^\infty \int_{\theta_\star}^\infty  \p(\theta_\star \given \vec{x},x_\star) \p(\theta_T \given x_T) \d\theta_T \d\theta_\star > 0.975 } \nonumber \\
&= \Infu{ \int_0^\infty  \p(\theta_\star \given \vec{x},x_\star)  \Pr(\theta_T>\theta_\star \given x_T)  \d\theta_\star > 0.975 } \label{eq:indicator}.
\end{align}
\changed{We base our calculations on being able to detect an increase of 0.12 in the true probability in the treatment arm over the control arm, therefore we set $\theta_T = \theta_\star + 0.12$.
To calculate the power for a given probability $\theta$, we take the expectation over all possible values over $x_\star$ and $x_T$:}
\begin{align*}
\text{Power}(\theta) = \sum_{x_T=0}^{n_T} \sum_{x_\star=0}^{n_\star} \p(x_T \given \theta+0.12) \p(x_\star \given \theta) \Ind{x_\star,x_T}.
\end{align*}

Once $\Ind{x_\star,x_T}$  in (\ref{eq:indicator}) has been calculated, it can also be used in the type I error calculations.

\paragraph{Type I error}
In regulatory contexts, type I error is perhaps the most important characteristic, as it represents incorrectly allowing an ineffective treatment to be approved. Bayesian methods with informative priors necessarily have some increase in type I error compared to standard frequentist methods. 
While a large increase in type I error is not acceptable, \citet{Grieve2016} argues that a small increase should not prevent the use of Bayesian methods.

Using the same test to determine difference as for power, we sum the results over all possible outcomes but assume the true parameter for the treatment and control are identical, \ie $\theta_T = \theta_\star$, therefore
\begin{align*}
\text{Type I error}(\theta) = \sum_{x_T=0}^{n_T} \sum_{x_\star=0}^{n_\star} \p(x_T \given \theta) \p(x_\star \given \theta) \Ind{x_\star,x_T}.
\end{align*}

\changed{
\paragraph{Pre-experimental rejection ratio}
The pre-experimental rejection ratio is calculated as
$$R_{\text{pre}}(\theta) = \frac{\text{Power}(\theta)}{\text{Type I error}(\theta)}.$$
\citet{bayarri_etal2016} have proposed to combine type I error and
power in this way in order to quantify the evidentiary impact of
statistical significance. This measure is particularly useful in our
setting, where both power and type I error depend on the true
parameter value $\theta$.  }

\section{Simulation Study}\label{sec:simulations}
\changed{
We conduct a simulation study similar to previous studies
\citep{viele2014use, Gravestock2016}, but with a few changes.  We now
have multiple historical studies, which are sampled from a random
effects model. We sample the historical data because we want to
understand how the methods perform on average, rather than for a
particular combination of studies. This is in contrast to the previous
simulation studies which had fixed historical data.  We use a uniform
$\Be(1,1)$ initial prior on the probabilities $\theta_\star$ and
$\theta_T$, which are the outcome probabilities in control and new
treatment arm, respectively.  
We consider all values of the treatment and control data, $X_T\in [0,n_T]$ and $X_\star\in [0,n_\star]$, which
are modelled as coming from the Binomial distributions $X_T\sim\Bin(n_T,\theta_T)$ and $X_T\sim\Bin(n_\star,\theta_\star)$, respectively.
The priors we consider are the NPP EB combined, EB Separate, and EB Pooled, the NPP FB with no correlation,
with correlation $\rho=0.5$ and the NPP FB pooled ($\approx
\rho=1$). For comparison we consider a MAP prior with
$\tau\sim\HN(1)$, the robustified MAP prior, and a prior with no
historical data (uniform on $\theta_\star$).  We conduct the
simulation study under two scenarios.
}

\paragraph{Scenario 1}
The first scenario considers a 5 small historical trials ($n_i=50$) being used to construct a prior for a larger new trial ($n_\star=200,n_T=200$) with 1:1 randomisation.
The historical data are sampled from distributions $X_i \sim \Bin(0.65+Z_i, 50)$ for $i= 1,..5$, where $Z_i \sim \Nor(0,0.1^2)$, with 1000 repetitions.

\paragraph{Scenario 2}
\changed{
The second scenario considers a 5 larger historical trials ($n_i=100$) being used to construct a prior for a smaller new trial ($n_\star=75,n_T=200$) with uneven randomisation to use fewer patients in the control arm.
The historical data are distributed $X_i \sim \Bin(0.65+Z_i, 100)$ for $i= 1,..5$, where $Z_i \sim \Nor(0,0.05^2)$, again with 1000 repetitions. 
}

For each iteration we calculate the operating characteristics for a
grid of $\theta$ values and then average these over the 1000
iterations.  \changed{Averaging the pre-experimental odds is based on the
geometric rather than the arithmetic mean.}
The simulation that considers a range of settings
represent biased and unbiased historical data. When the true parameter
is near the historical data, there is no bias and so this matches very
closely to the assumptions of the MAP model.  However, as difference
between the true parameter and the historical data increases, so does
the bias. Therefore this simulation design tests a broad range of bias
and therefore model suitability.  The results are shown in two groups: 
the first compares the EB methods with no borrowing, and the second 
contrasts the EB Combined with the full Bayes power priors and MAP priors.

\subsection{Prior Sample Size}\label{sec:PSS}

\begin{figure}[htbp]
	\begin{minipage}{0.5\textwidth}
  	\includegraphics[width=\textwidth]{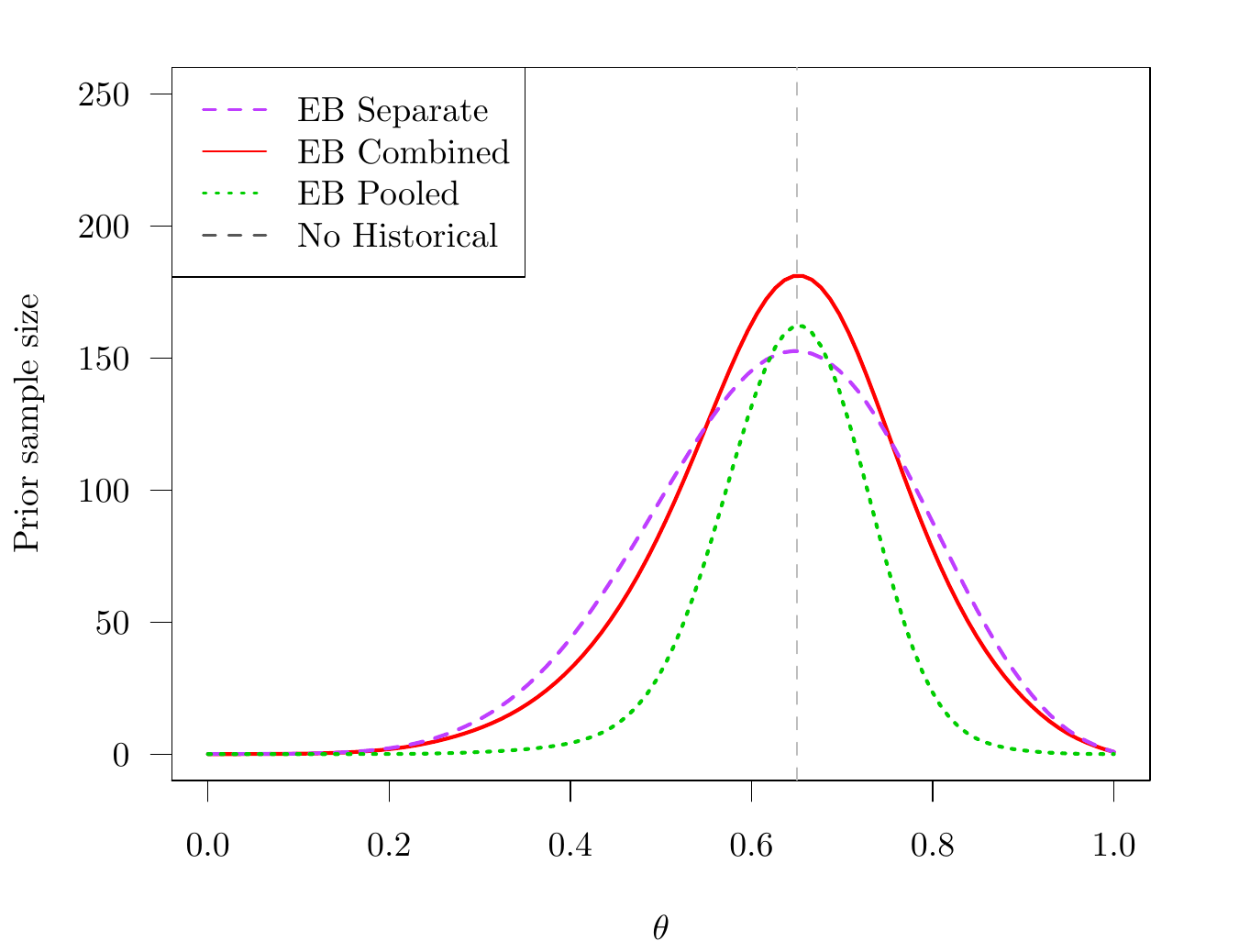}
	\end{minipage}
	\hfill
	\begin{minipage}{0.5\textwidth}
  	\includegraphics[width=\textwidth]{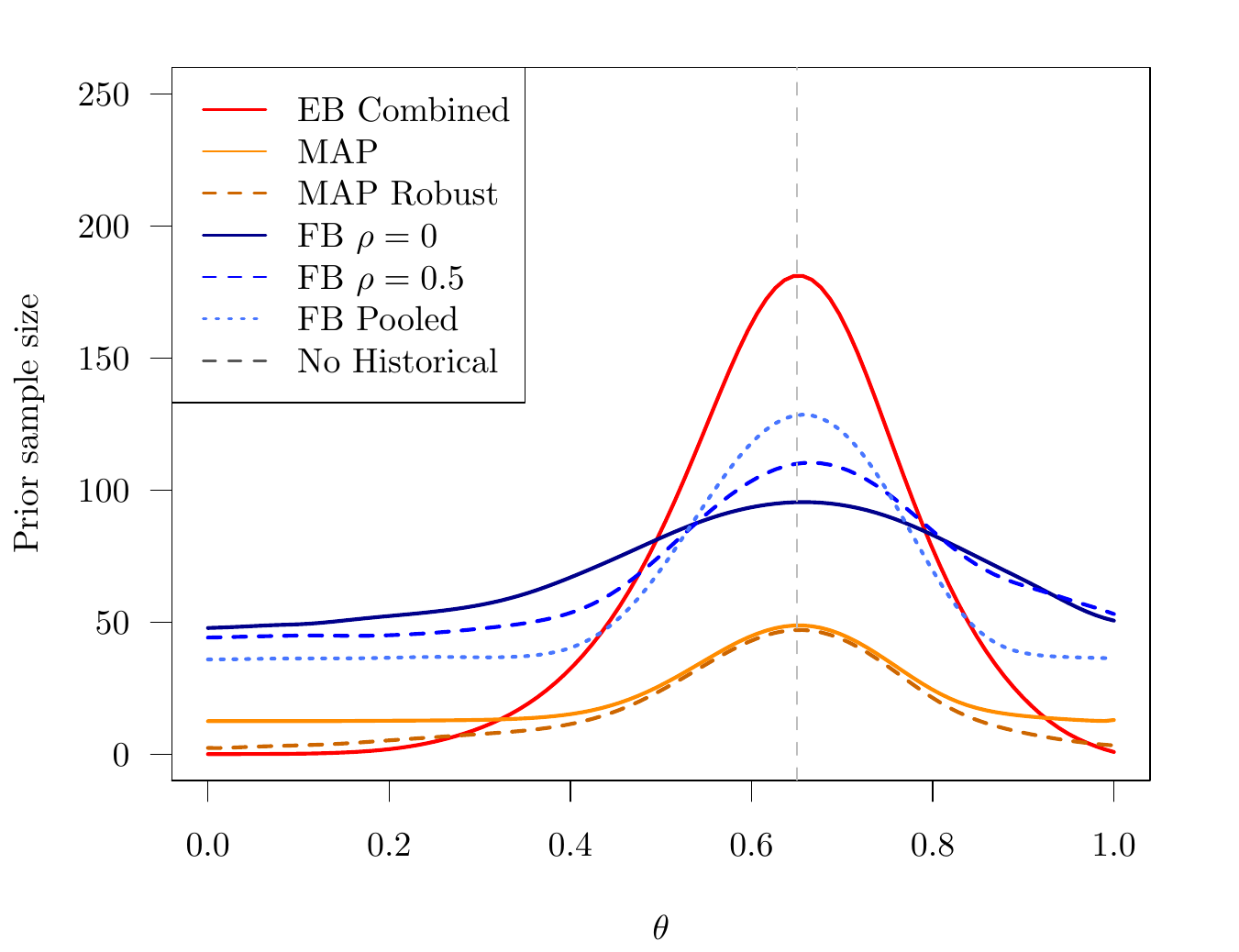}
  \end{minipage}
	
		\begin{minipage}{0.5\textwidth}
  	\includegraphics[width=\textwidth]{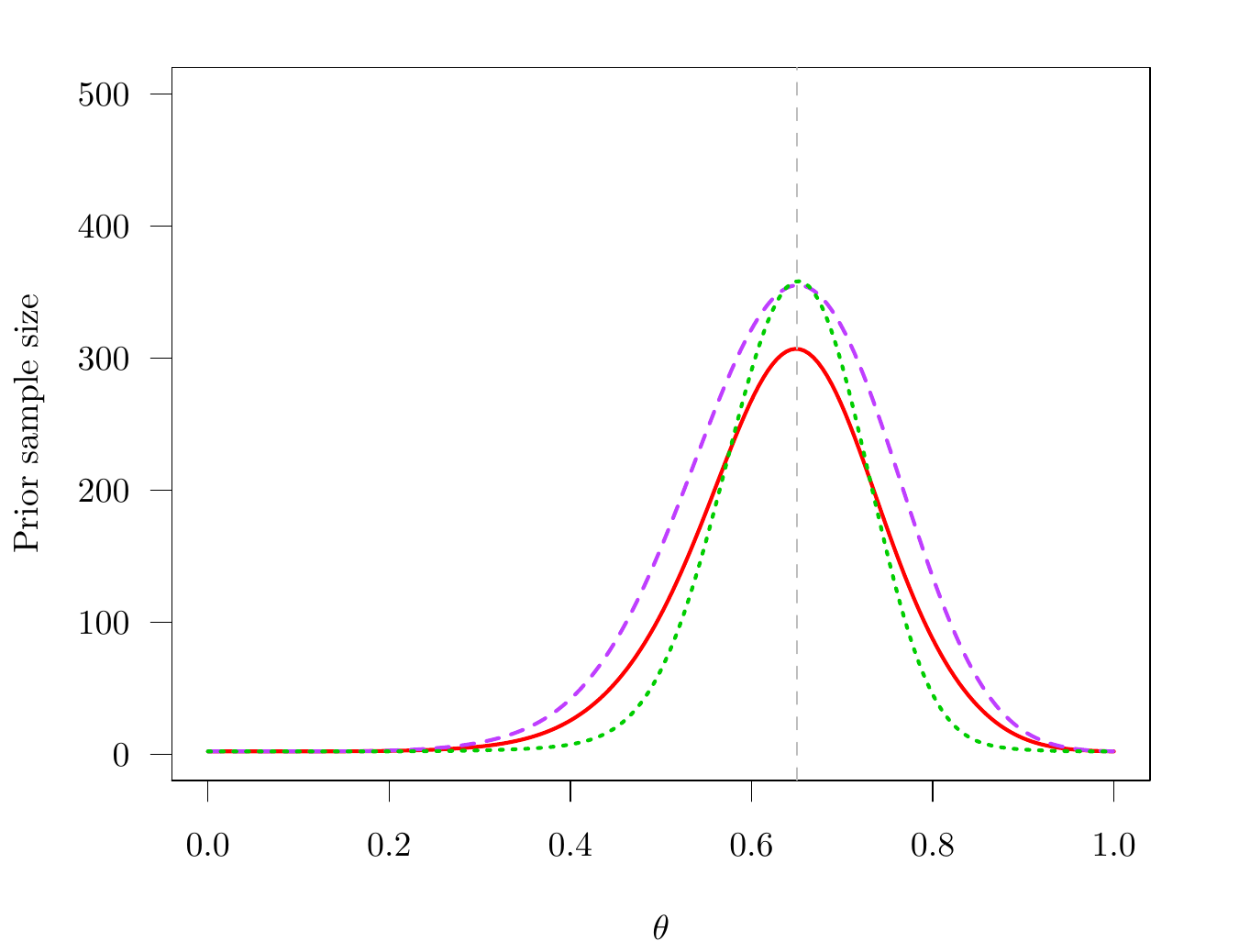}
	\end{minipage}
	\hfill
	\begin{minipage}{0.5\textwidth}
  	\includegraphics[width=\textwidth]{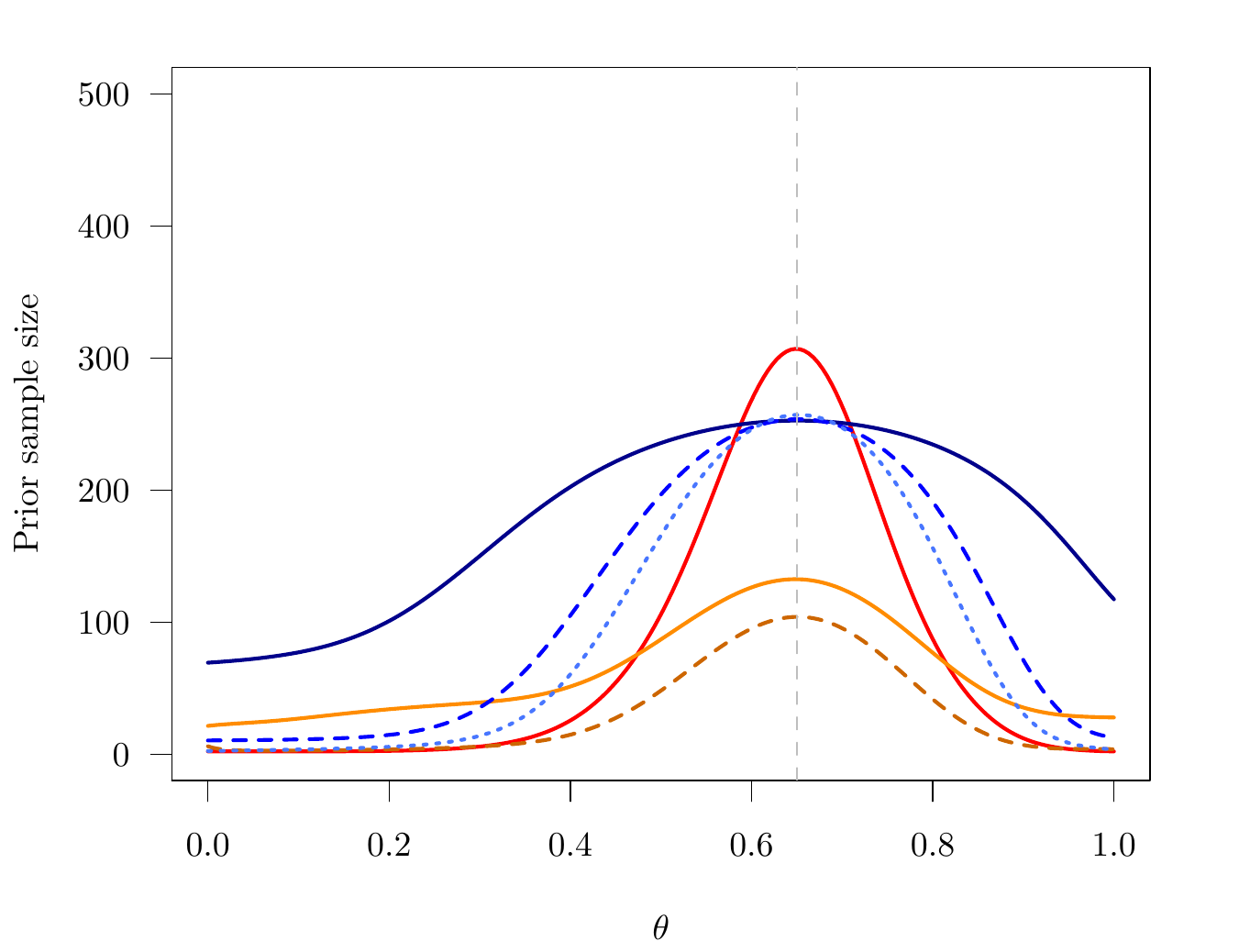}
  \end{minipage}
	
	\caption{Prior sample sizes for the EB priors (left) and FB and MAP priors (right) for scenarios 1 (above) and 2 (below).}
  \label{fig:PSS}
\end{figure}

\changed{
The method of constructing the prior and the choice of hyper-parameters strongly
influences how much information is incorporated into the prior and thus how strongly
the prior influences the posterior. By quantifying the amount of historical
information included in the prior, we can explain much of the behaviour of the
other operating characteristics of the methods. Figure \ref{fig:PSS} shows the 
expected sample sizes of the priors in Scenario 1 and 2.}

\changed{
The plots of the left column of Figure \ref{fig:PSS} show the sample sizes for the
three EB methods for choosing $\hat{\vec{\delta}}$. The EB Pooled has
the narrowest range of borrowing, which is due to the borrowing only occurring
when new data is close to the single large ``pooled" trial. 
Wider is EB Combined, which borrows from a combination of trials when
the new data is within the range of the historical studies, otherwise it mostly 
borrows from the most extreme trial only. EB Separate borrows over the same range as EB Combined,
but has larger sample size in the tails, because it often gives non-zero weight to 
the next closest study, where EB Combined has 0, as seen in the example in Section
\ref{sec:antibiotics}, Table \ref{tab:estimates}.
}

\changed{
The FB power priors have broader prior sample size curves than the EB priors. 
This is partially due to them being fully Bayesian, and thus incorporating additional
uncertainty compared to the EB methods, but also due to them being fixed with regard
to the new data. There is a parallel between the behaviour of the EB Pooled and Separate
priors and the FB pooled and FB $\rho=0$ priors. The FB Pooled borrows over a narrower
range, but has a slightly larger sample size when the current data and the historical
data align, while FB $\rho=0$ borrows the most over the widest range. This can be
explained by to the correlation parameter, where by assuming the $\delta_i$ are
independent more information is taken from each $x_i$ to estimate them. 
The MAP prior includes considerably less information when the historical data aligns
with the new data than the other priors. 
Notably, all of these fully Bayesian priors' sample size does not drop to 0 in the tails.
This difference can be directly seen by comparing the MAP and MAP robust in the two plots.
Due to the fatter tails, these priors will have larger impacts on the posterior
than the adaptive methods, EB and MAP robust. 
A benefit of this property is that the adaptive priors will have less of an influence
away from the historical information.
}

\subsection {Comparison of Other Operating Characteristics}

\changed{For all of the different priors, the same general pattern can be
observed in the operating characteristics. The relative advantages and disadvantages
of the prior construction methods depend on which operating characteristics is
most relevant to the required analysis. We describe the operating characteristics
in terms of the true parameter in the control arm, $\theta$.}

\changed{
The MSE is generally low where the historical data and $\theta$ align
and increases as $\theta$ differs. Most of the priors are not too strong in the tails
and the MSE begins to drop again for large and small $\theta$ values.}

The type I error is low for $\theta$ values lower than the historical data and 
increases for $\theta$ larger. All of the methods have some increase in type I
error relative to including no historical information in the prior. This is 
due to the prior biasing the posterior towards the historical data and thus making
an incorrect significant result more or less likely.

\changed{
The power shows the other side of the trade-off of including historical data. 
When $\theta$ is smaller than the historical data, the control arm is biased
up relative to the truth and there is a smaller chance to declare a result significant
and therefore a loss of power. For $\theta$ above the historical data, there is an 
increase in power relative to including no historical data.}

\changed{
The pre-experimental rejection ratio summarises the trade-off between power and 
type I error, and therefore is larger for smaller values of $\theta$ where type I
error is small and decreases as $\theta$ and the type I error increases.
The change in the ratio is mostly influenced by the type I error as its relative
change with $\theta$ is much larger than that of power. Generally, we see that the 
priors have a larger rejection ratio than the prior without historical data for small
values of $\theta$ and have lower rejection ratios for large values. This suggests 
there is some disadvantage in all of the priors examined for testing purposes.
The important consideration is how much this ratio changes and how much lower is it
than the no historical prior.
}

\changed{
As suggested in Section \ref{sec:PSS}, the differences in the operating characteristics
is due to the strength of the priors, which can be explained through the prior sample size.
The other contributing factor is due to the different scenarios. Scenario 1 has
much more new data relative to historical data than Scenario 2. Therefore the
priors in Scenario 1 are much less able to overwhelm the likelihood of the new data
and thus the operating characteristics are more stable.}

\begin{figure}[htbp]

  \begin{minipage}{0.5\textwidth}
\end{minipage}
	
	\begin{minipage}{0.5\textwidth}
	  \includegraphics[width=\textwidth]{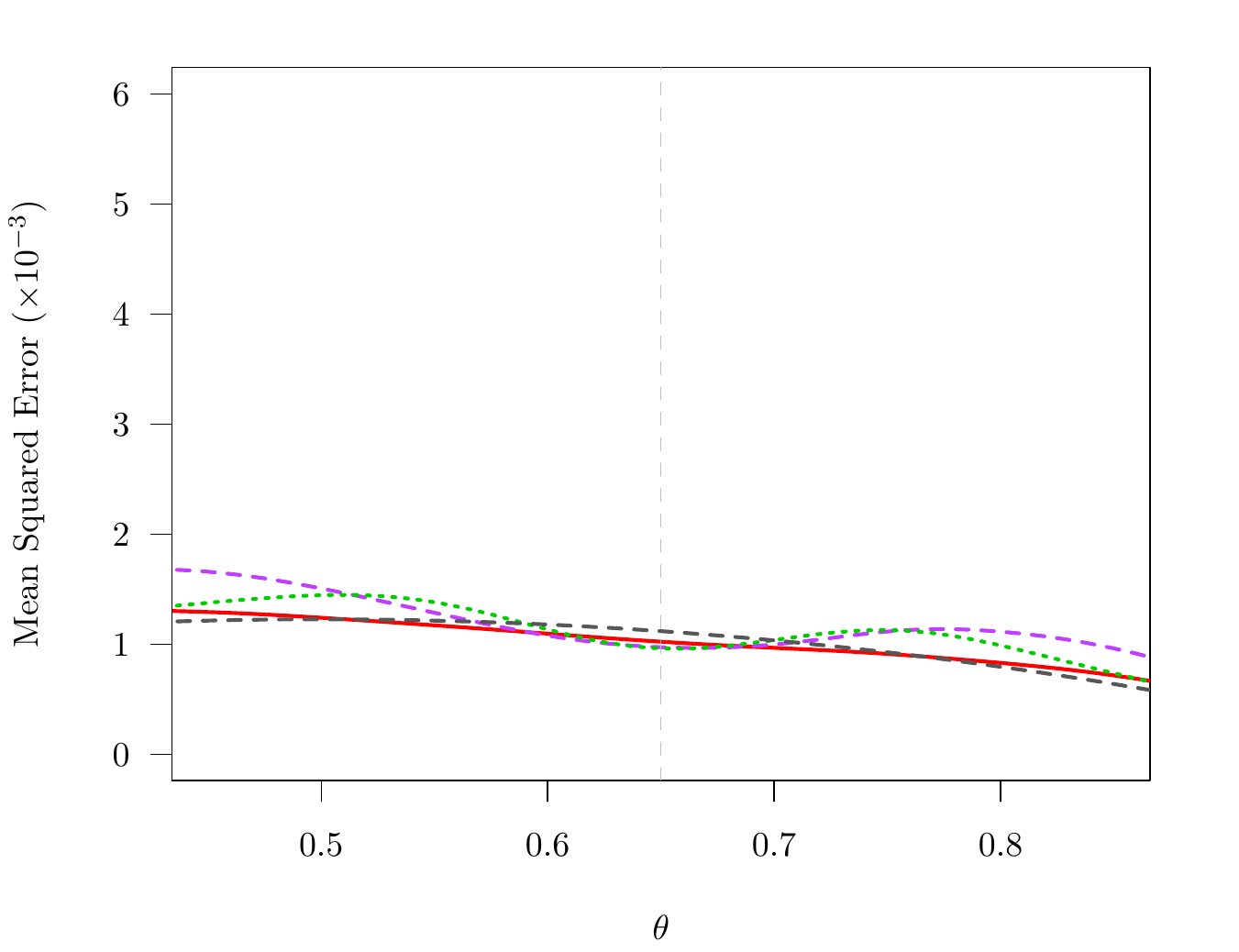}
	\end{minipage}
	\hfill
	\begin{minipage}{0.5\textwidth}
  	\includegraphics[width=\textwidth]{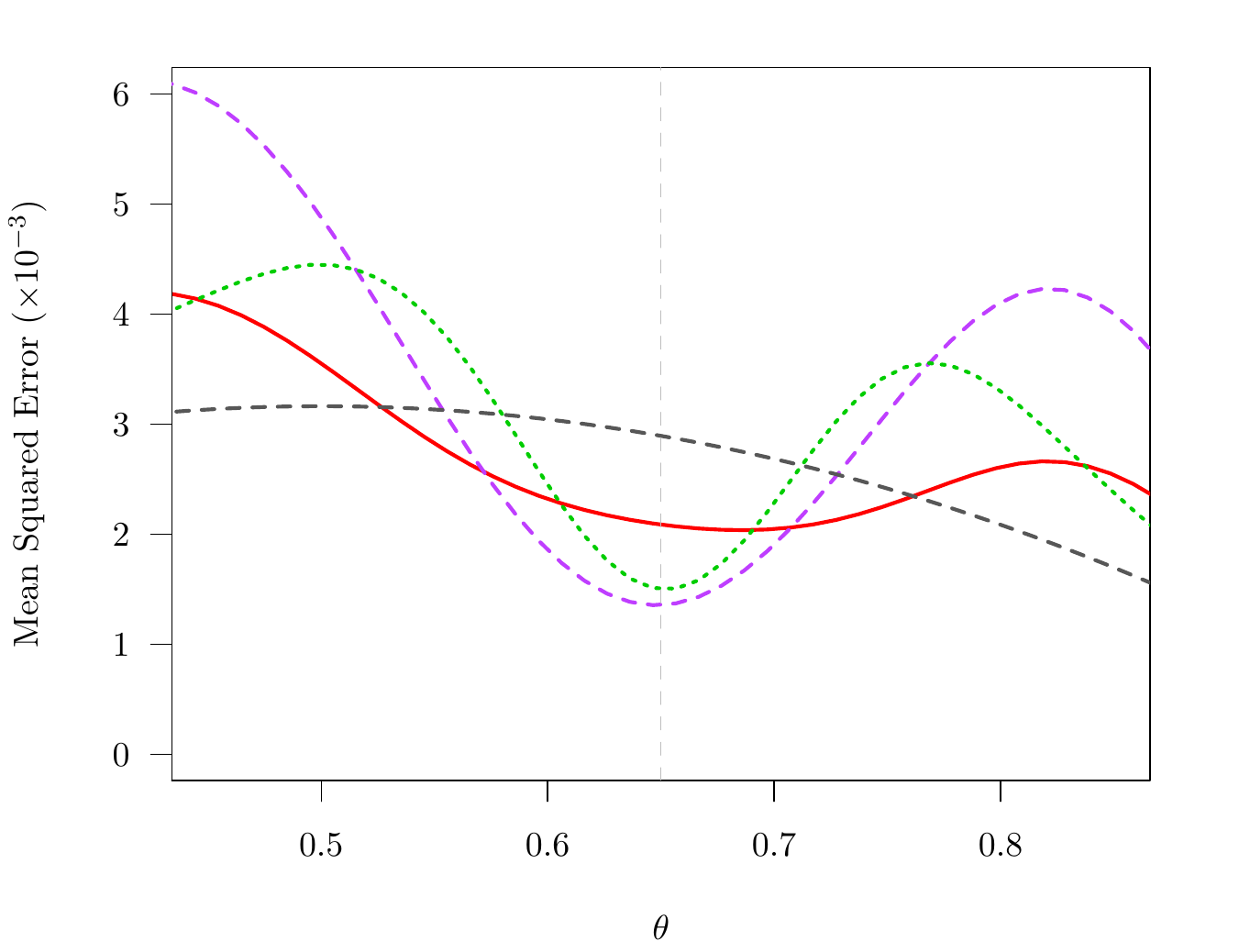}
	\end{minipage}

	\begin{minipage}{0.5\textwidth}
	  \includegraphics[width=\textwidth]{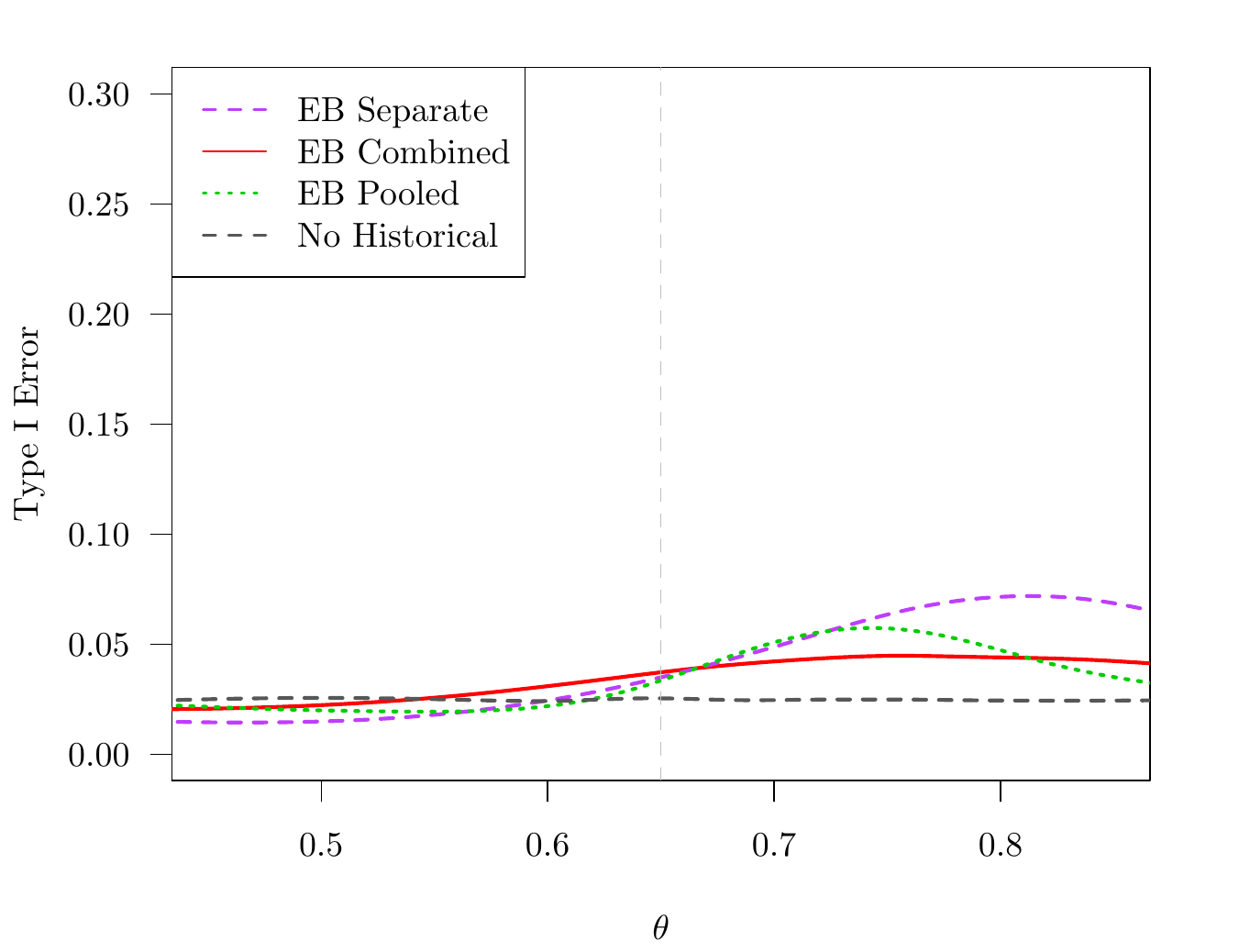}
	\end{minipage}
	\hfill
	\begin{minipage}{0.5\textwidth}
	  \includegraphics[width=\textwidth]{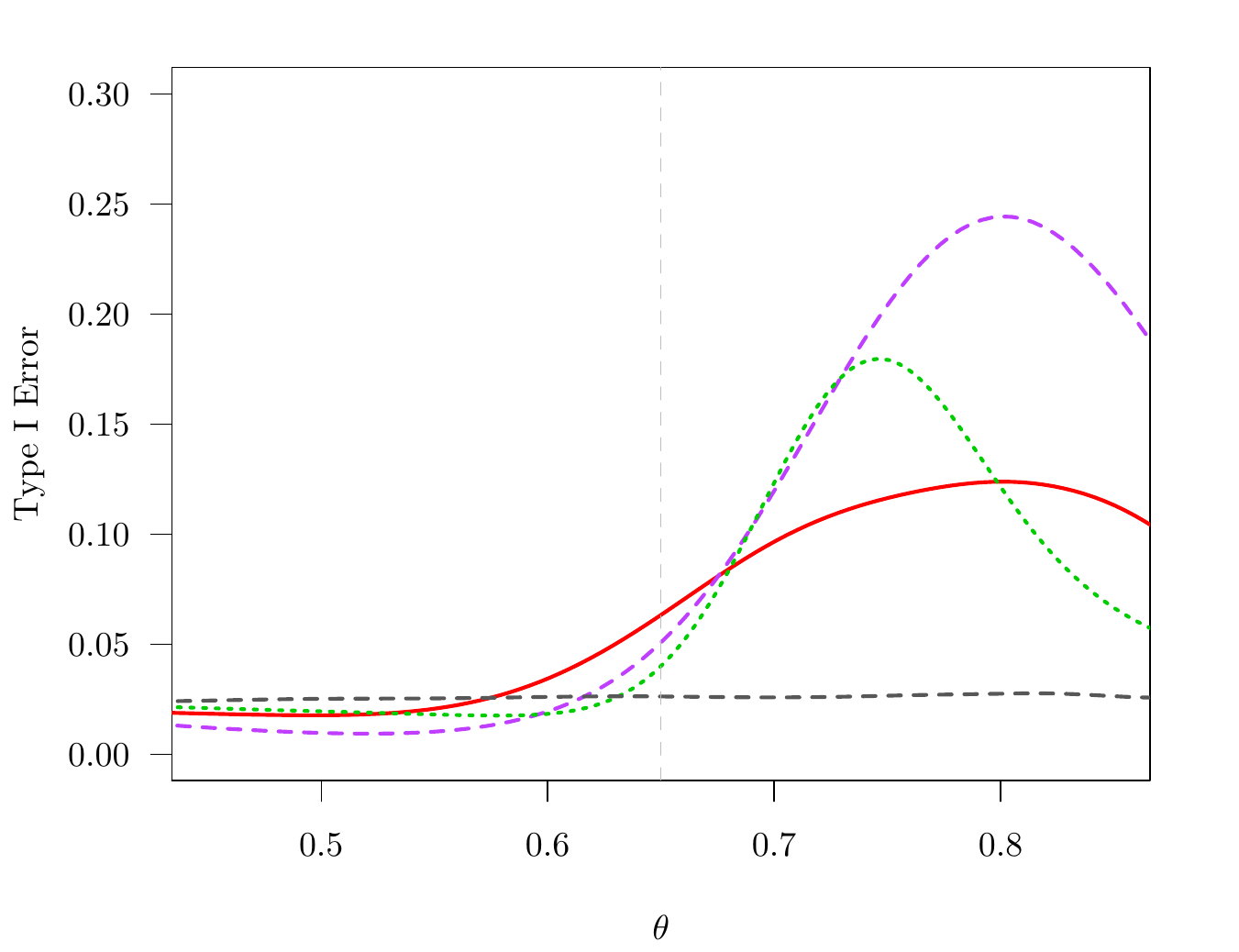}
	\end{minipage}

	\begin{minipage}{0.5\textwidth}
	  \includegraphics[width=\textwidth]{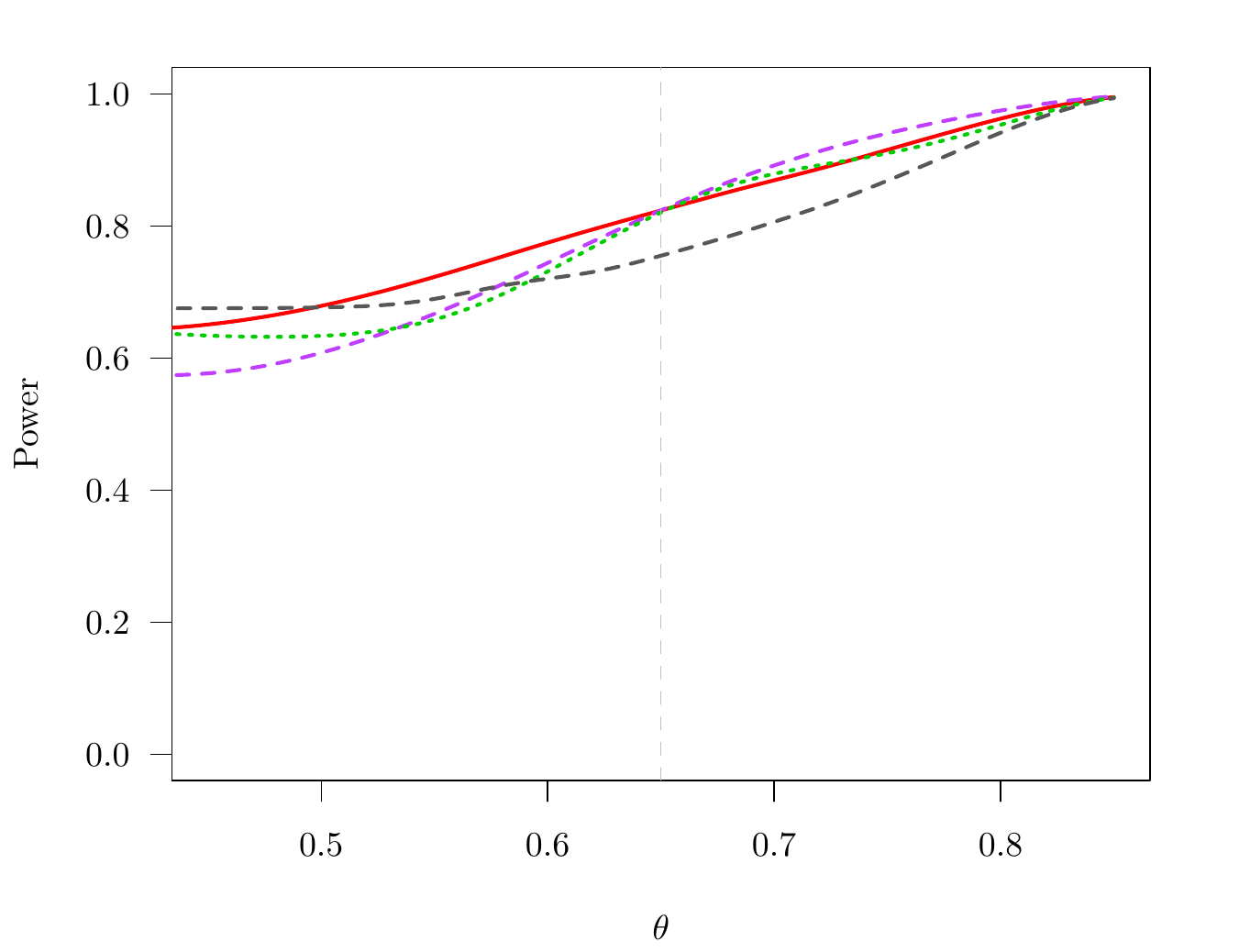}
	\end{minipage}
	\hfill
	\begin{minipage}{0.5\textwidth}
  	\includegraphics[width=\textwidth]{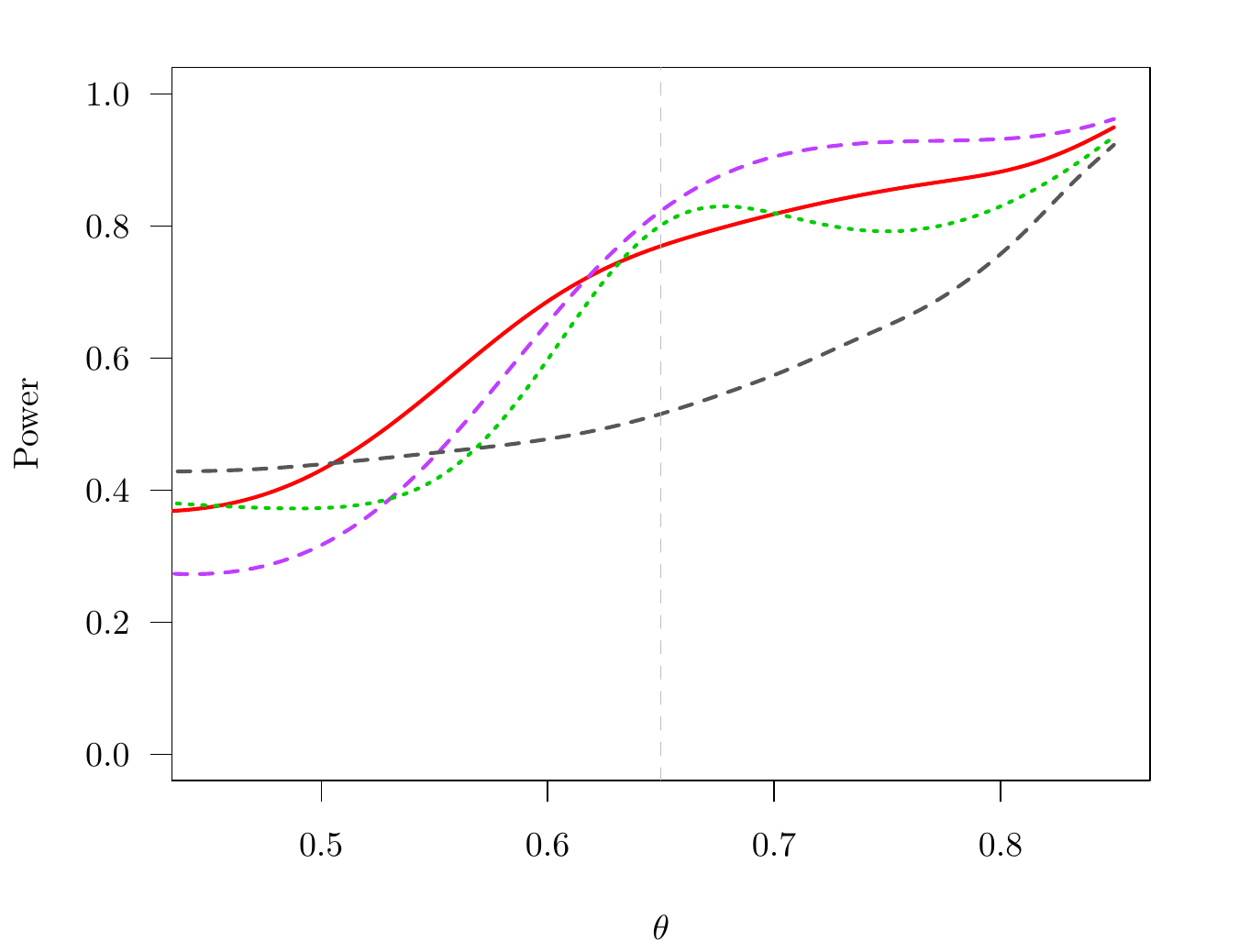}
	\end{minipage}

		  \begin{minipage}{0.5\textwidth}
	  \includegraphics[width=\textwidth]{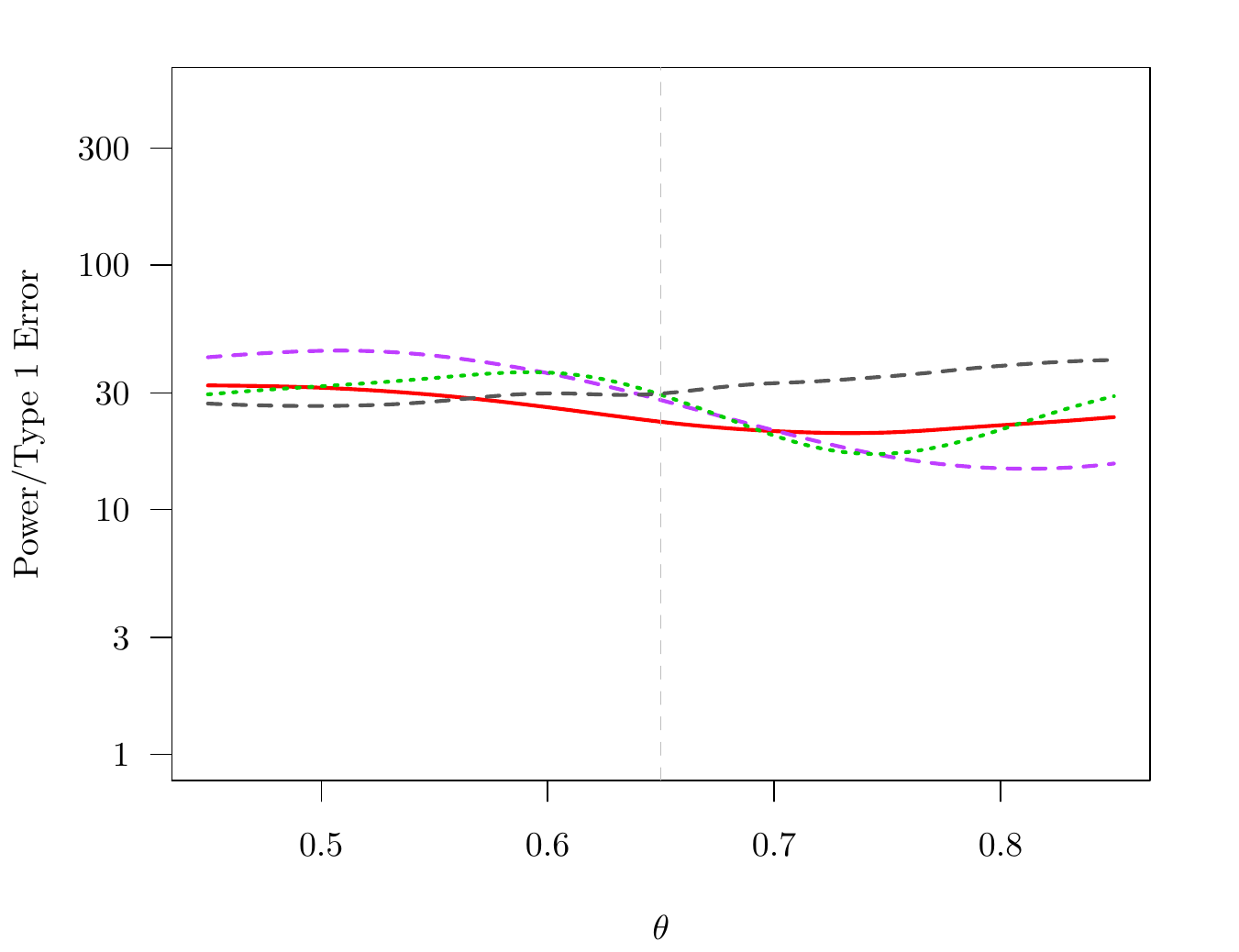}
  \end{minipage}
  \hfill
  \begin{minipage}{0.5\textwidth}
	  \includegraphics[width=\textwidth]{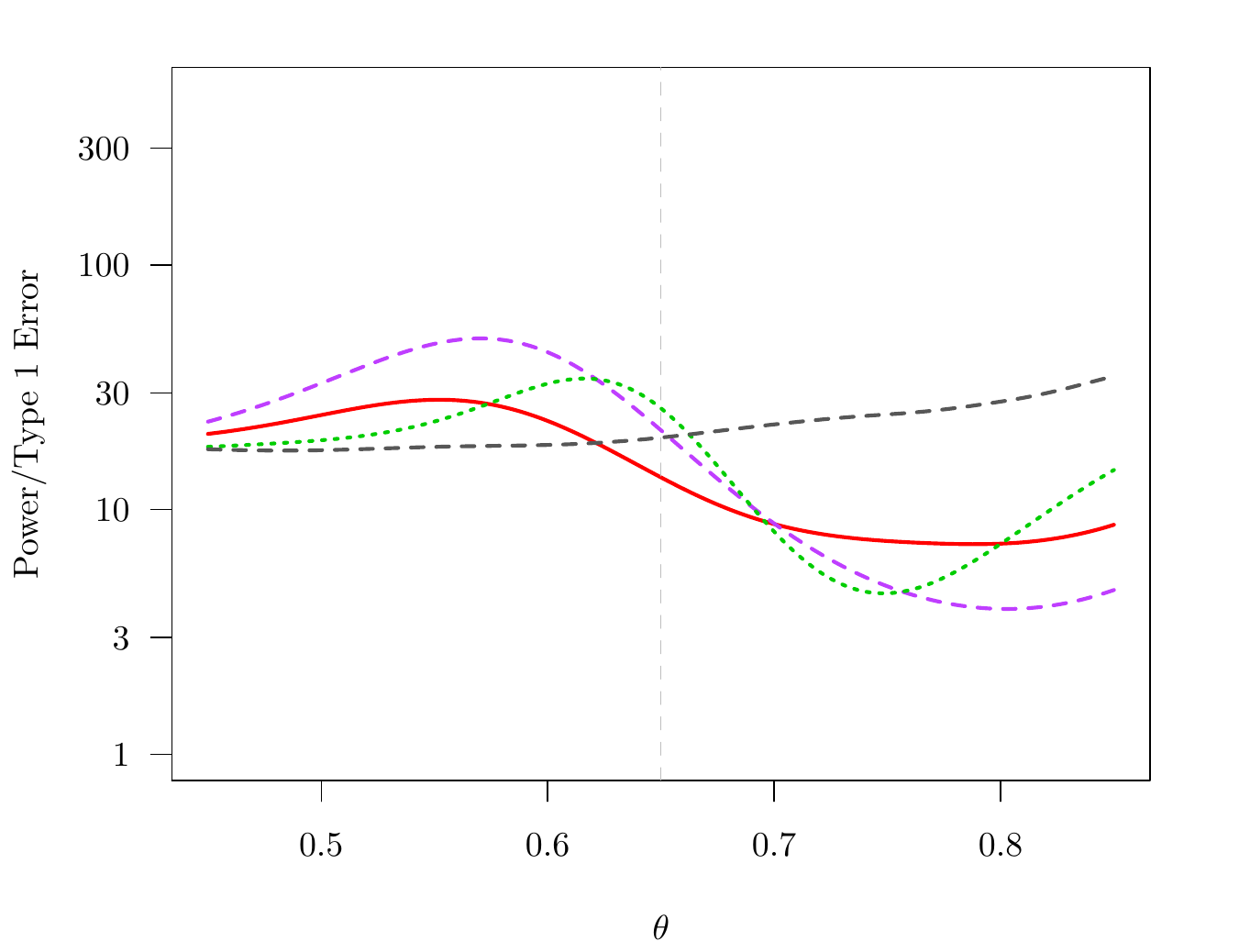}
  \end{minipage}

 \caption{Operating characteristics of the posteriors based on EB type priors for true parameter $\theta$ for scenarios 1 (left column) and 2 (right column).}
 \label{fig:OCEB}
\end{figure}

\changed{
Figure \ref{fig:OCEB} shows the operating characteristics for the EB PP methods.
The differences between methods are not very pronounced in Scenario 1, but are 
more obvious in Scenario 2. The relative performance among the priors is the same in
both Scenarios. Following the general patterns described previously, EB Separate 
has the most extreme behaviour, having largest MSE increase and the largest change 
in type I error and power, which corresponds to the wide range of and high level of borrowing.
This is summarised in the rejection ration, where EB Separate has the largest and the smallest values.
The narrow borrowing of EB Pooled can also been seen in the operating characteristics,
where the changes occur in a narrower band of $\theta$ values than for EB Combined and 
EB Separate.
The operating characteristics of EB Combined are flatter than the other two EB methods,
having less MSE, less loss of power, less type I error increase and, correspondingly, a 
more stable rejection ratio. This flatness is a desirable quality as it means the outcome
of the inference is less dependent on the true value of $\theta$.
}

\begin{figure}[htbp]

  \begin{minipage}{0.5\textwidth}
\end{minipage}
	
	\begin{minipage}{0.5\textwidth}
	  \includegraphics[width=\textwidth]{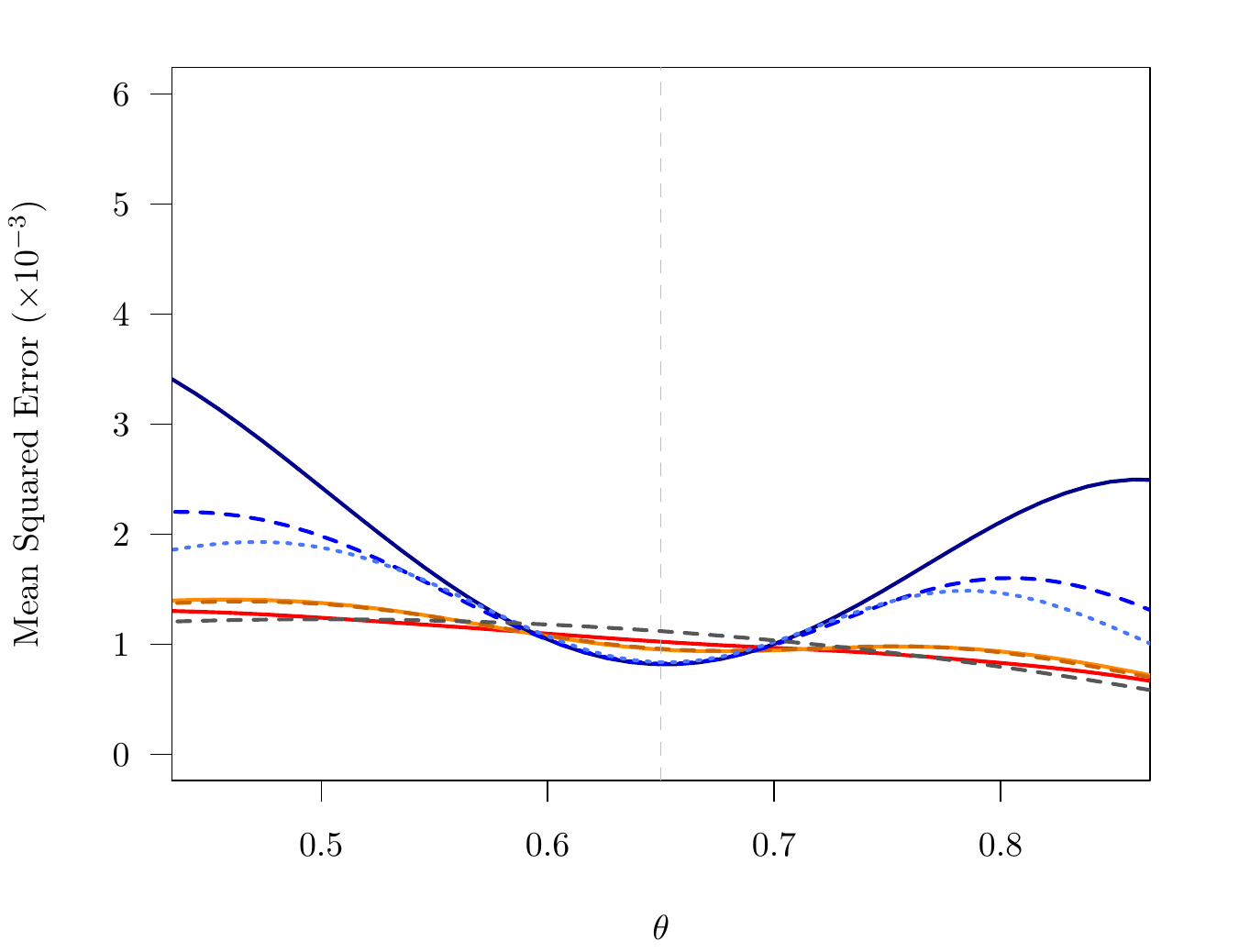}
	\end{minipage}
	\hfill
	\begin{minipage}{0.5\textwidth}
  	\includegraphics[width=\textwidth]{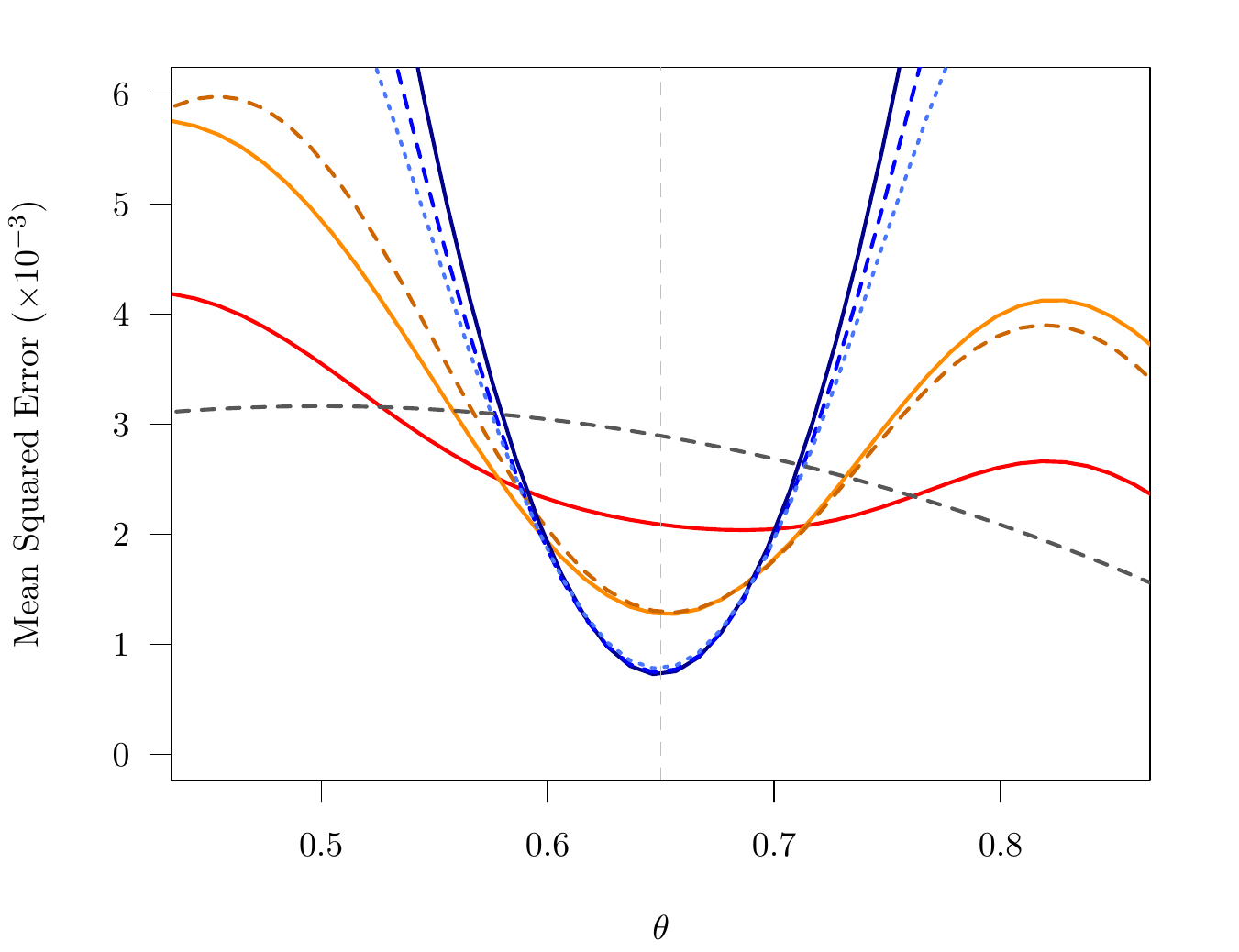}
	\end{minipage}

	\begin{minipage}{0.5\textwidth}
	  \includegraphics[width=\textwidth]{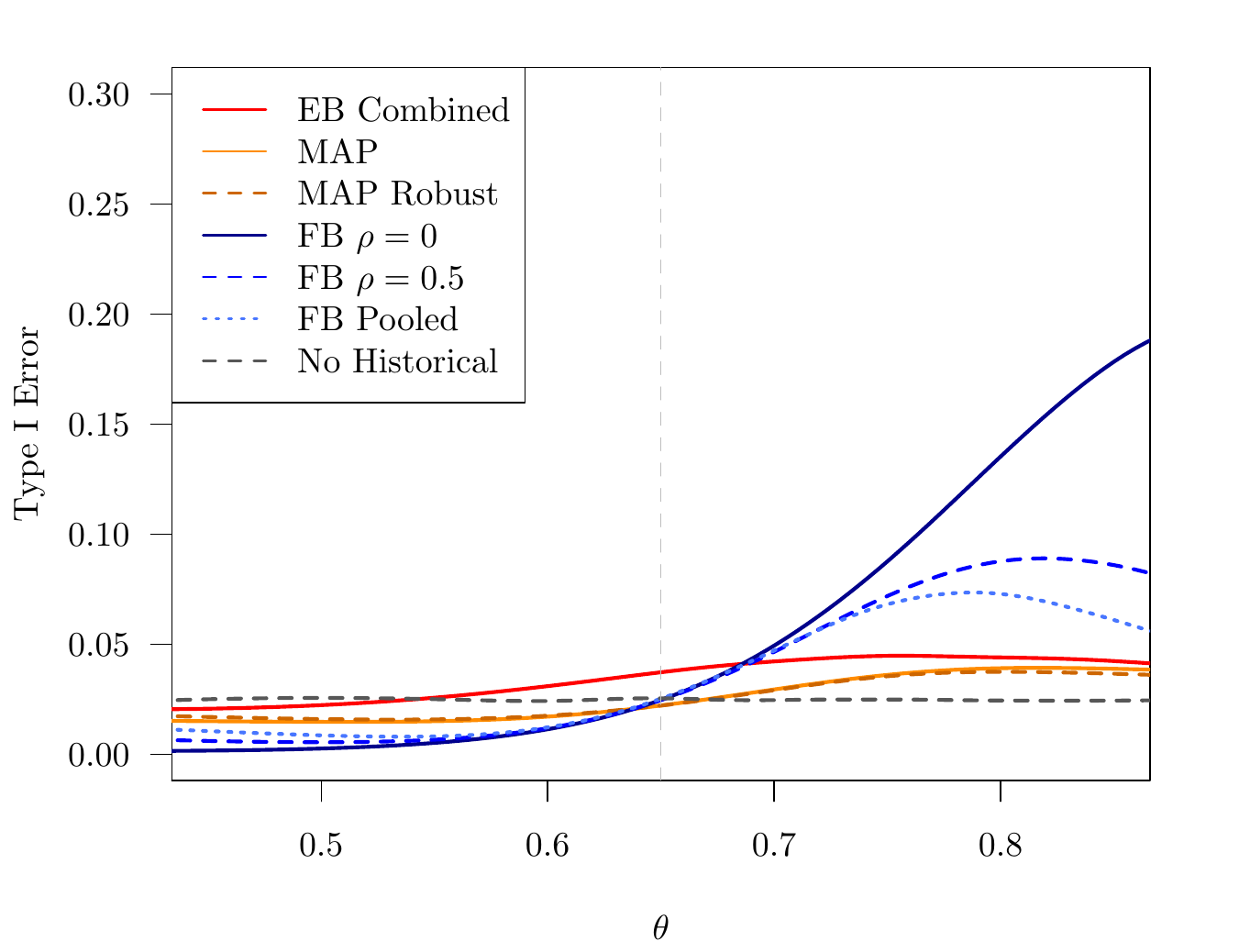}
	\end{minipage}
	\hfill
	\begin{minipage}{0.5\textwidth}
	  \includegraphics[width=\textwidth]{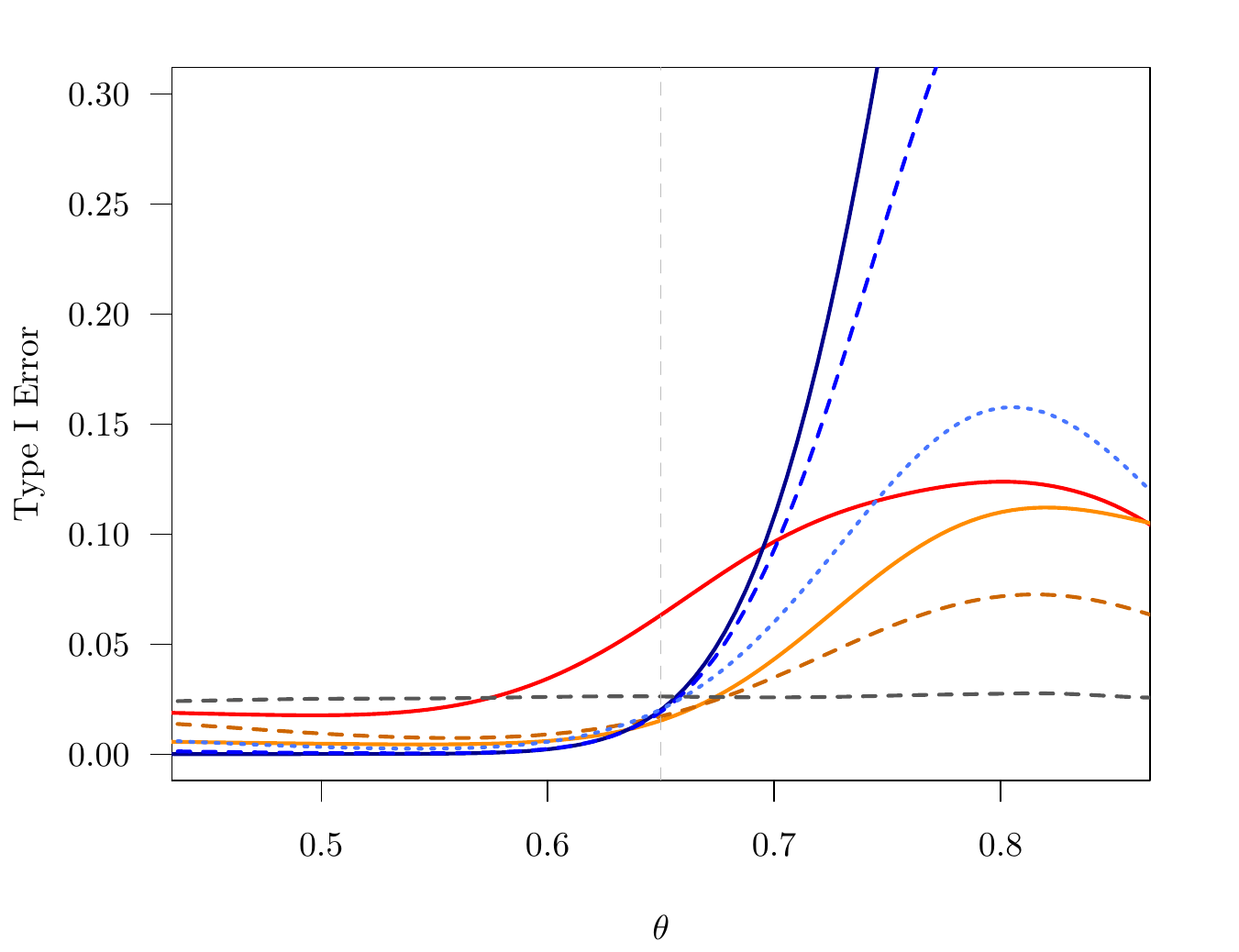}
	\end{minipage}

	\begin{minipage}{0.5\textwidth}
	  \includegraphics[width=\textwidth]{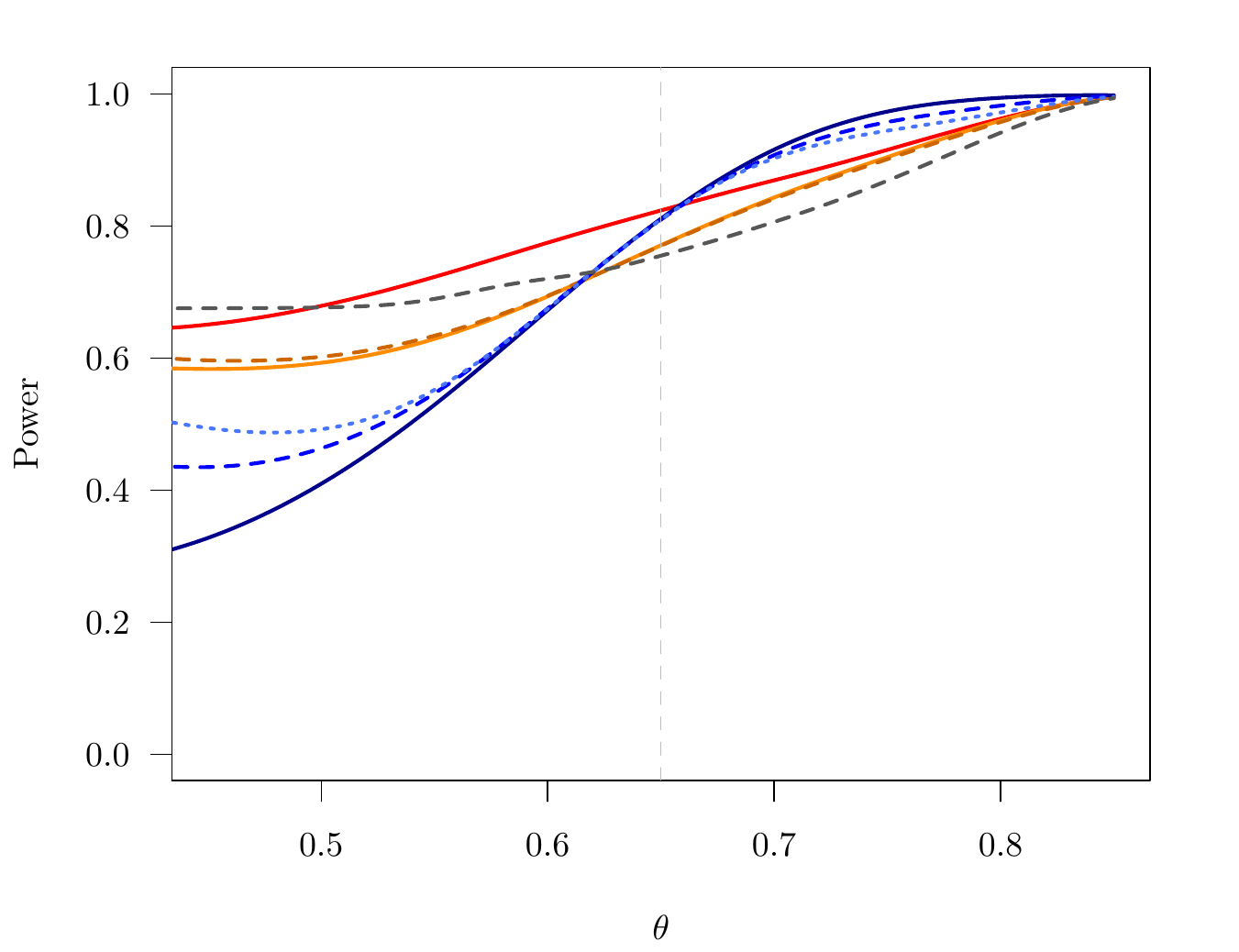}
	\end{minipage}
	\hfill
	\begin{minipage}{0.5\textwidth}
  	\includegraphics[width=\textwidth]{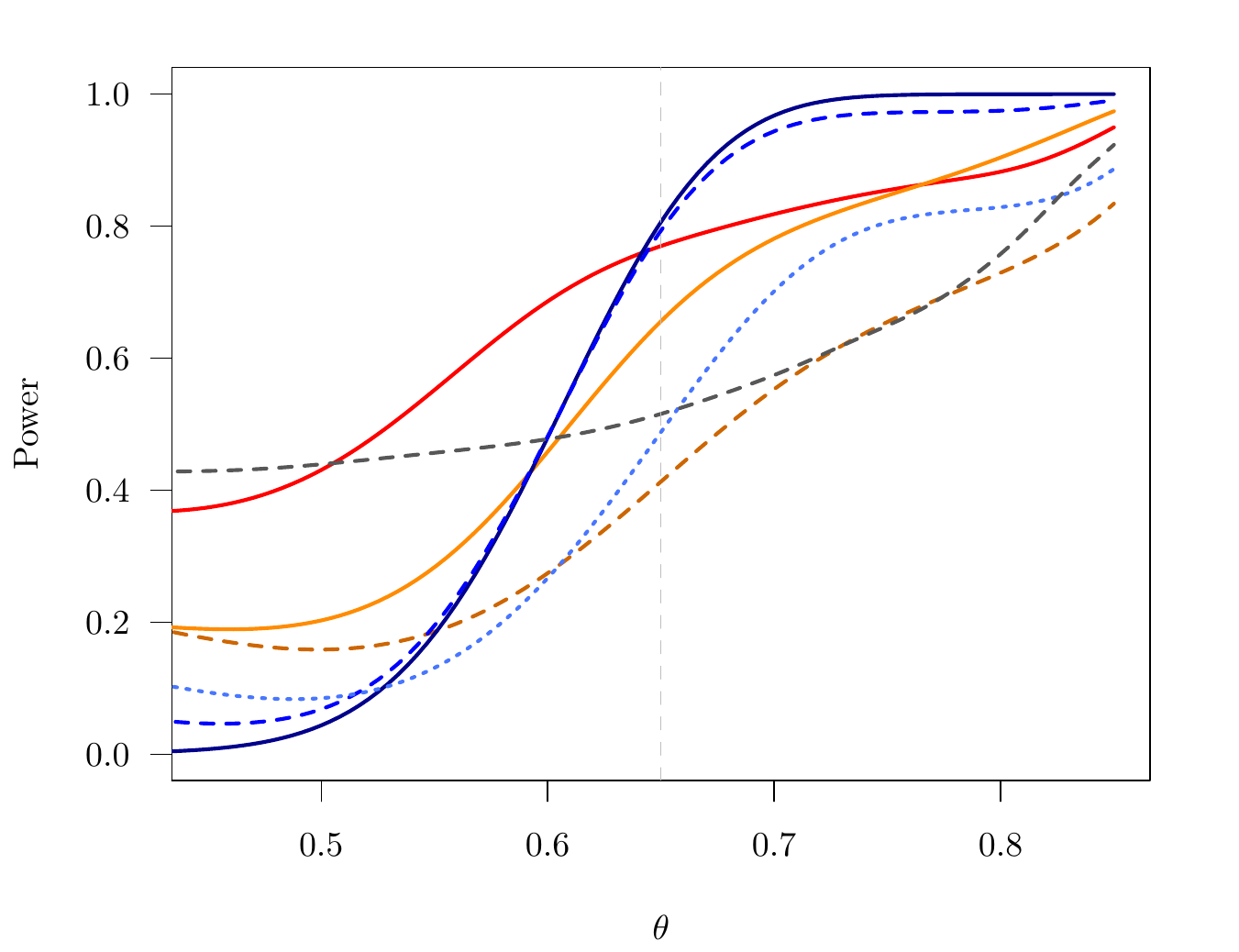}
	\end{minipage}

  \begin{minipage}{0.5\textwidth}
	  \includegraphics[width=\textwidth]{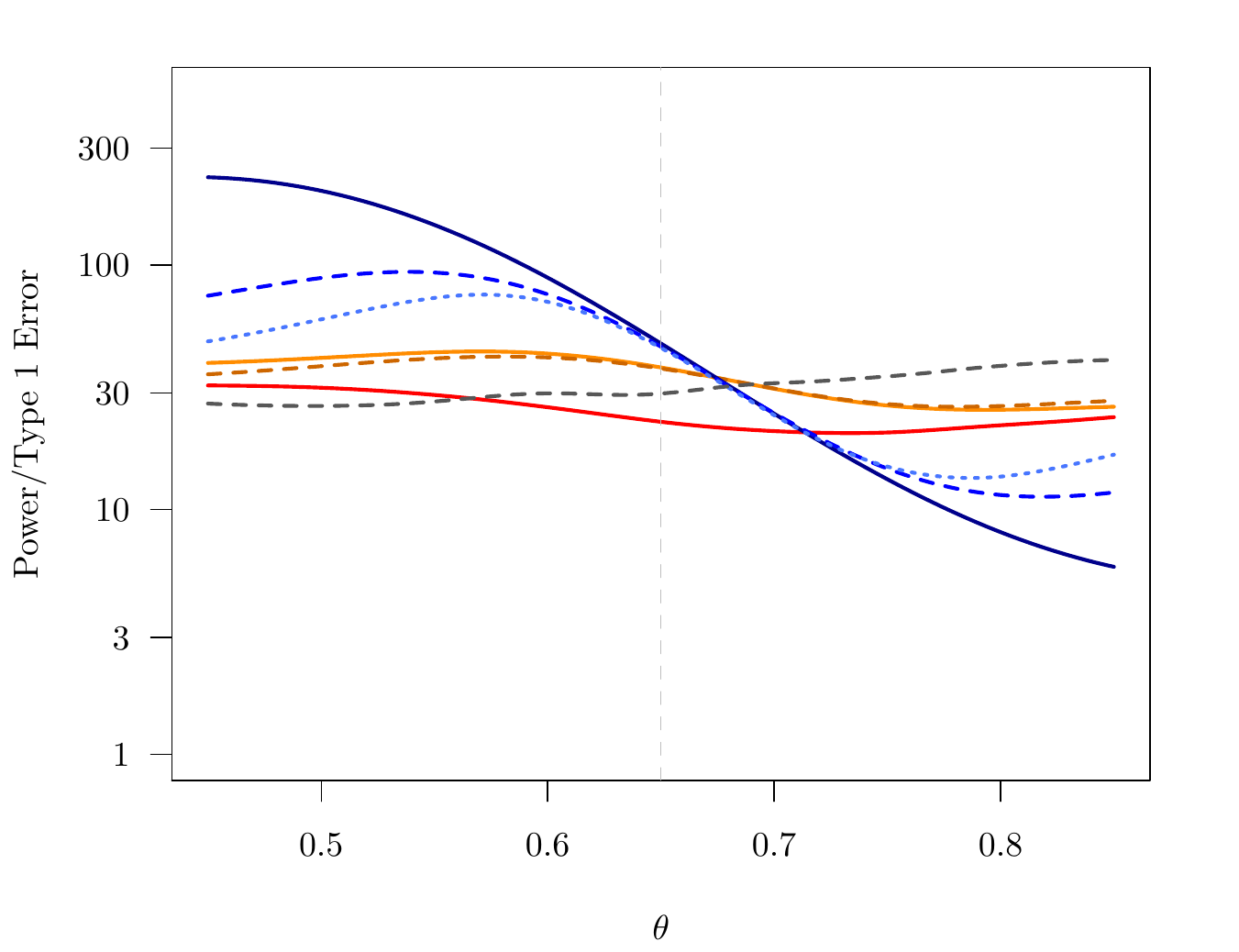}
  \end{minipage}
  \hfill
  \begin{minipage}{0.5\textwidth}
	  \includegraphics[width=\textwidth]{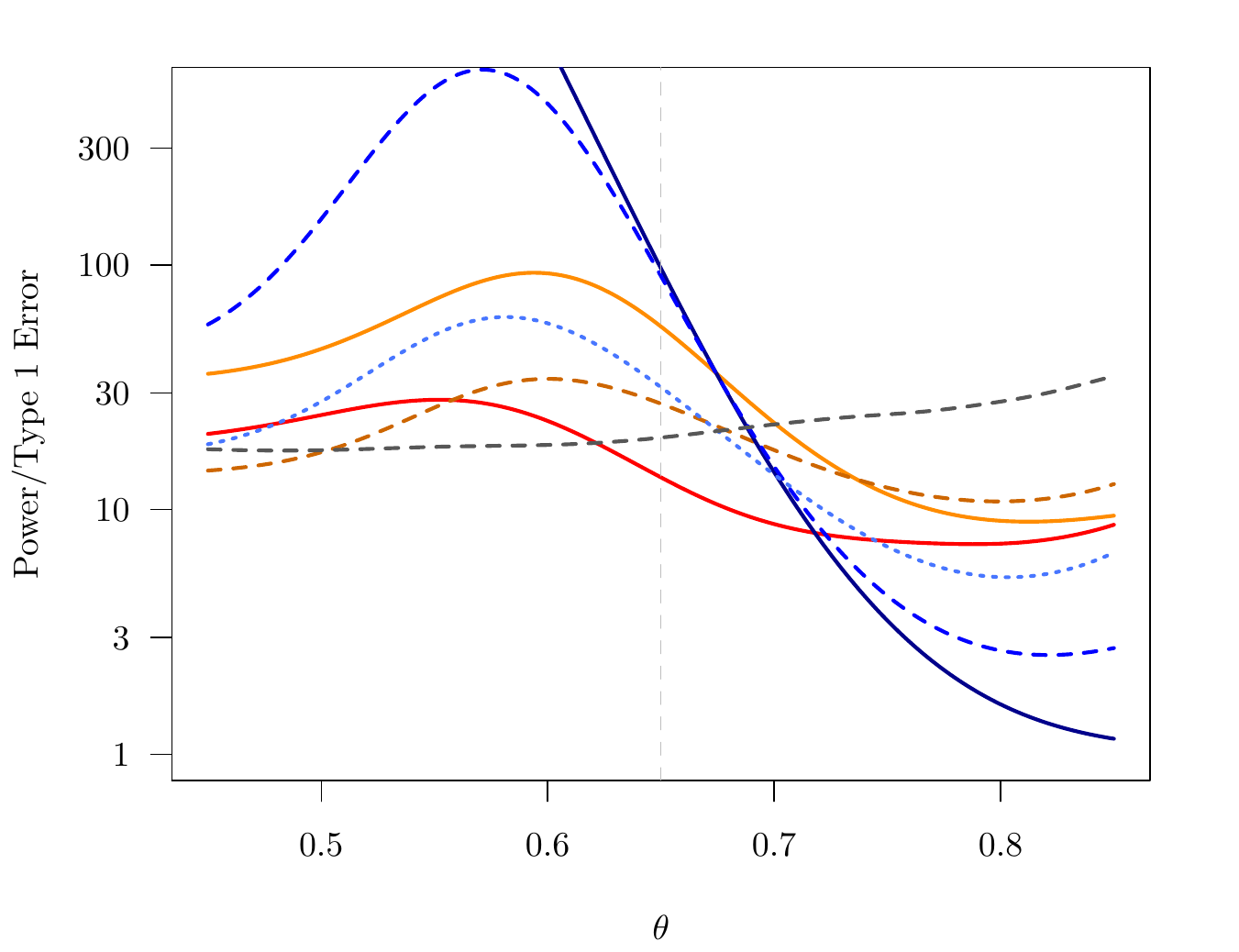}
  \end{minipage}

 \caption{Operating characteristics of the posteriors based on FB PP and MAP methods for true parameter $\theta$ for scenarios 1 (left column) and 2 (right column).}
 \label{fig:OCFB}
\end{figure}

\changed{
Figure \ref{fig:OCFB} shows the operating characteristics for the FB PP and MAP methods,
with EB Combined shown for reference. The three FB priors, pooled, $\rho=0.5$ and $\rho=0$,
are clearly related and the plots show that the correlation parameter $\rho$ is effective
in adjusting performance. With $\rho=0$ the operating characteristics vary wildly, with
large MSE, large type I error increase, large power loss, and huge change in rejection ratio etc. This can be explained by
its large prior sample size which does not decrease in the tails and therefore biases the
posterior very strongly when $\theta$ is small or large. For $\rho=0.5$, with less prior
sample size and weaker prior, the changes in operating characteristics are slightly smaller,
and for FB Pooled, \ie $\rho=1$, the operating characteristics are much more stable.
Notably for FB Pooled in Scenario 2, the rejection ratio behaves more closely to MAP and
EB Combined than to the other FB priors.
}

\changed{
The MAP prior has much less pronounced changes in operating characteristics due to the 
correspondingly small prior sample size. The benefit of this limited borrowing is that
the type I error increase is limited and there is a less extreme loss of power than 
the FB PP methods. The robust MAP is even more conservative in borrowing, especially
in Scenario 2. This prevents type I error increase and reduces MSE, but has a negative
impact on the power. Indeed for all values of $\theta$ it has less power than using 
no historical data at all.
}

\changed{
Relative to the MAP and FF priors, the EB Combined increases in power and type I error
for smaller values of $\theta$. However, based on the pre-experimental rejection ratio
the performance is quite stable, and therefore the prior's effect is more reliable.
}

\section{Discussion}\label{sec:discussion}
This paper describes a new approach to estimating the parameters of a power prior
when used with multiple historical studies.
The EB power prior methodology provides a data-driven approach to construct 
informative priors based on historical trial data.
It is computationally cheap and the simulation study shows it has
good operating characteristics. A very convenient feature is the simple
formula for sample size, which is important in applications. 
Considering the naïve EB type approaches, the EB Pooled approach ignores the
variability of the studies.
The EB Separate approach does not perform well compared to the EB Combined method which
is based on the proper distribution, and so despite the simple interpretation of
the weights as compatibility measures for each study, it cannot be advocated.
In the EB Combined approach, the estimated weights $\vec{\hat\delta}_{\text{EB}}$ 
give a combination of the historical data that is most compatible with the new
data. Therefore, weights less than 1 can be interpreted as penalising
studies that make the sum of the historical evidence incompatible with
the newly observed data.
FDA guidance for device trials \citep{BayesianFDA2010} recommends methods to prevent
borrowing in case of conflict and the EB Combined method has the best operating characteristics
of the EB methods and also is the most principled from a mathematical perspective.

\changed{
In this paper, we also compare the operating characteristics of meta-analytic 
and other fully Bayesian power prior methods for constructing priors based on
multiple historical studies.
The MAP and its robust version perform very well in our tests, but the performance
of the EB approach has some positive characteristics, which,
depending on the requirements of a study design could be used to good advantage.}
Surprisingly, our simulations show that the fully Bayesian approach to the power prior
with a flat prior on the weights is a poor choice for multiple
studies, while the EB approach is much better.
These results are in contrast to the results of \citet{Gravestock2016}
where, for a single historical study, the full Bayes and empirical Bayes methods
had similar performance and neither was clearly better.

\changed{
The simulation results of the FB methods suggest that there is not
information available to override the prior used for the weight parameters.
Judged by its poor operating characteristics, the default multi-variate uniform prior is a poor choice.
With the connection made in Section \ref{sec:normal} between power priors and meta-analytic models, 
it might be possible to consider priors used in meta-analysis and transform them for the power prior weights.
The correlated prior for FB brings the power prior closer to the exchangeability model, in the sense that the historical studies are used together, which increases the information used in determining the distribution of the weight parameters, making the prior more stable.
}

Although our studies have been mostly limited to binomial setting,
we expect the empirical Bayes estimates to have similar behaviour for other likelihoods:
giving weight to the most similar model if the new data is outside the range of
the historical studies, and when within the range, a combination of weights 
that make the historical data similar to the new data.

An essential assumption of these methods is that the historical study
data used are suitable to estimate the new study parameter.  It
is the responsibility of the practitioner to evaluate the studies and
decide if they should be included in the prior. 
For the power prior, based on the ``equal but discounted" assumption,
the following question must be asked: Is this historical data informative about the parameter of interest?
The methods presented here attempt to find the best way to generate the prior given the
data, they have no capacity to determine if a study is not suitable.
The methods do not necessarily break down if the studies chosen are
unsuitable, but the inference based on the priors may become
unjustifiable.


\paragraph{Acknowledgements}
We thank Beat Neuenschwander and two referees for helpful comments on a previous version of this manuscript.

\appendix


 \bibliography{ms}
\end{document}